\documentclass[reprint,twocolumn,aps,pre,floatfix,amssymb,superscriptaddress]{revtex4-1}
\usepackage[utf8]{inputenc}
\usepackage{amssymb,bm,xspace,soul} 
\usepackage{color} 
\usepackage[papersize={8.5in,11in}]{geometry}
\usepackage{natbib}
\usepackage{graphicx}
\usepackage{amsmath}
\usepackage[table]{xcolor}
\usepackage{subcaption}
\usepackage{verbatim}
\usepackage{braket}
\usepackage{enumitem}
\usepackage{math tools} 
\usepackage{dsfont} 
\usepackage{mathrsfs} 
\usepackage{multirow}
\usepackage{slashed}

\usepackage{setspace}

\usepackage{subcaption}
\usepackage[justification=raggedright]{caption}
\usepackage{pifont}

\usepackage[colorlinks=true]{hyperref} 
\hypersetup{
    bookmarks=true,         
    unicode=false,          
    pdftoolbar=true,        
    pdfmenubar=true,        
    pdffitwindow=false,     
    pdfstartview={FitH},    
    pdftitle={My title},    
    pdfauthor={Author},     
    pdfsubject={Subject},   
    pdfcreator={Creator},   
    pdfproducer={Producer}, 
    pdfkeywords={keyword1} {key2} {key3}, 
    pdfnewwindow=true,      
    colorlinks=true,       
    linkcolor=magenta, 
    citecolor=blue,        
    filecolor=magenta,      
    urlcolor=cyan           
} 

\makeatletter
    \def\CT@@do@color{%
      \global\let\CT@do@color\relax
            \@tempdima\wd\z@
            \advance\@tempdima\@tempdimb
            \advance\@tempdima\@tempdimc
    \advance\@tempdimb\tabcolsep
    \advance\@tempdimc\tabcolsep
    \advance\@tempdima2\tabcolsep
            \kern-\@tempdimb
            \leaders\vrule
                    \hskip\@tempdima\@plus  1fill
            \kern-\@tempdimc
            \hskip-\wd\z@ \@plus -1fill }
    \makeatother

\renewcommand{\P}{\mathcal{P}}

\newcommand{\vn}{{\boldsymbol{n}}}

\newcommand{\vN}{{\boldsymbol{N}}}

\renewcommand{\o}{\over}
\newcommand{\eq}[1]{\begin{align}#1\end{align}}

\renewcommand{\(}{\left(}
\renewcommand{\)}{\right)}
\renewcommand{\[}{\left[}
\renewcommand{\]}{\right]}
\newcommand{\abs}[1]{\left| #1 \right|}

\newcommand{\nt}{\notag\\}

\let\v\boldsymbol

\newcommand{\ep}{\epsilon}

\renewcommand{\a}{\alpha}

\renewcommand{\d}{\delta}
\newcommand{\g}{\gamma}

\newcommand{\s}{\sigma}

\renewcommand{\th}{\theta}

\newcommand{\lam}{\lambda}

\renewcommand{\dag}{\dagger}

\newcommand{\Zt}{\mathds{Z}_2}

\usepackage{tocloft} 
\addtocontents{toc}{\cftpagenumbersoff{chapter}}
\addtocontents{toc}{\cftpagenumbersoff{section}}

\setlist[enumerate]{leftmargin=*}

\DeclareMathAlphabet{\mathcalb}{U}{BOONDOX-calo}{m}{n}
\SetMathAlphabet{\mathcalb}{bold}{U}{BOONDOX-calo}{b}{n}
\DeclareMathAlphabet{\mathcalbb}{U}{BOONDOX-calo}{b}{n}

\DeclareMathAlphabet{\mathcald}{U}{dutchcal}{m}{n}
\SetMathAlphabet{\mathcald}{bold}{U}{dutchcal}{b}{n}
\DeclareMathAlphabet{\mathcaldb}{U}{dutchcal}{b}{n}

\DeclareFontFamily{OT1}{pzc}{}
\DeclareFontShape{OT1}{pzc}{m}{it}{<-> s * [1.10] pzcmi7t}{}
\DeclareMathAlphabet{\mathpzc}{OT1}{pzc}{m}{it}

\renewcommand{\S}{\boldsymbol{S}}

\newcommand{\cJ}{J}

\newcommand{\ttot}{{\mathrm{tot}}}
\newcommand{\vcS}{{\mathcalbb{S}}}
\newcommand{\etal}{\emph{et al.} }

\usepackage{todonotes}
\presetkeys{todonotes}{color=green!40,bordercolor=green,inline}{}

\begin{document}

\title{Simulating spin systems with Majorana networks}
 \author{Alex Thomson}
 \author{Falko Pientka}
  \affiliation{Department of Physics, Harvard University, Cambridge, Massachusetts, 02138, USA}
\date{\today}

\begin{abstract}
With the discovery of Majorana quasiparticles in semiconductor-superconductor hybrid structures, topologically protected qubits have emerged as a promising contender for quantum information processing. 
While the construction of a universal quantum computer with topological protection likely requires significant advances in materials science, intermediate-scale devices are nearly within the reach of current technology. 
As a near-term milestone for topological qubits, we propose a network of topological superconductors as a simulator of a large variety of quantum spin systems, including those with frustration.
Our proposal is founded on existing technology, combining advantages of semiconducting and superconducting qubits. 
We identify local measurement protocols that give access to information about ground and excited states as well as dynamic correlations. 
The topological protection of the qubits results in longer coherence times, and relaxation to the groundstate can be controlled by coupling the network an external bath. 
We conclude by pointing out specific applications of the quantum simulator, \emph{e.g.}, spin liquids, quantum criticality, and thermalization.

\end{abstract}

\maketitle

\renewcommand{\arraystretch}{1.3}

\section{Introduction}
The prospect of quantum computing has made the field of quantum information one of the most rapidly developing areas in physics.
Nevertheless, in spite of encouraging results with up to 20 qubits \cite{Friis2018,ibmq}, the realization of a full-scale quantum computer might still be in the distant future. The fundamental challenge is to maintain coherence of a large number of qubits \cite{Linke2017}. Protecting quantum information against noise requires the implementation of error correction schemes, but this comes at the cost of a large overhead, with the number of physical qubits typically exceeding the number of logical qubits by several orders of magnitude \cite{Campbell2017}.

The need for error correction can be reduced by using topological qubits formed by multiple spatially-separated Majorana quasiparticles living at domain walls of one-dimensional topological superconductors. The information is stored nonlocally and is therefore immune to local low-frequency noise. While a large variety of topological-superconductor platforms are being investigated in the lab \cite{Mourik2012,Albrecht2016,Nadj-Perge2014,Ruby2015,Nichele2017,He2017,Liu2018,Hart2014,Hart2016,Pientka2017,Hell2017,Lee2017}, the most encouraging results have been achieved in semiconductor devices with proximity-induced superconductivity \cite{Lutchyn2018}.
This includes the initial discovery of Majorana states \cite{Mourik2012} and the subsequent observation of unique Majorana signatures such as an exponential protection \cite{Albrecht2016}, a quantized conductance \cite{Zhang2018}, and a fractional Josephson effect \cite{Laroche2017}.
Motivated by these findings,  several detailed proposals for quantum computers consisting of arrays of mesoscopic superconducting islands hosting Majorana states \cite{Fu2010,Heck2012,Hyart2013,Karzig2017,Plugge2017,Vijay2016}
have been made recently \cite{Vijay2015,Landau2016,Litinski2017}. 

The efficient simulation of quantum many-body systems is a prime motivation for building a computer.
While the latter remains unavailable, quantum simulation has been demonstrated in a variety of systems throughout the last decade \cite{Bloch2012,Blatt2012,Aspuru-Guzik2012,Houck2012,quantumSimReview,quSimUltraCold}. The underlying idea of quantum simulation is to carefully tune the microscopic parameters of a many-body system so that it mimics the behavior of another many-body system on certain time scales. 
A key advantage of analogue quantum simulators is that far fewer resources are required relative to a quantum computer since, ideally, every physical qubit corresponds to a logical qubit.
Nevertheless, quantum simulators are equally plagued by decoherence for physically relevant system sizes. Moreover, they require delicate control over parameters.

In this paper, we propose the realization of an analogue topological quantum simulator of spin systems. We envision a network of coupled superconducting islands hosting four Majorana states, each representing a spin-$1/2$ site \cite{Beri2012,Altland2013}. By suitably arranging the islands, one can simulate a plethora of quantum spin models, including models on a variety of lattices with spin greater than or equal to $1/2$.
The exponential protection of Majorana states results in long coherence times while the flexibility of the semiconductor platform allows for intricate and versatile control.
Moreover, the charging energy in mesoscopic superconducting islands suppresses quasiparticle poisoning, one of the main sources of error of Majorana qubits.

Motivated by the experimental effort to realize a quantum computer in semiconducting hybrid structures \cite{Lutchyn2018}, we focus on an implementation of Majorana qubits in a semiconductor platform. 
By combining well-established methods of manipulating superconducting and semiconducting qubits in the lab, we propose realistic setups to measure and control the simulated spin system. 
These include coupling our system to transmission line resonators and techniques developed for the control of quantum dots.

We remark that our quantum simulator is similar in spirit to certain quantum computer proposals, {\it e.g.}, the surface code realizations of Refs.~\citenum{Vijay2015} and \citenum{Landau2016}. Moreover, the simulator requires similar hardware components as the topological quantum computer proposed, {\it e.g.}, in Refs.~\citenum{Karzig2017,Plugge2017,Vijay2016}. Our quantum simulator can therefore serve as a important milestone in the effort towards the long-term goal of building a topological quantum computer.

This paper is organized as follows. Section~\ref{sec:setup} presents the general setup of the spin-1/2 island and how it can be used to generate spin models. Section~\ref{sec:ExpRealization} proposes a specific experimental setup using semiconductor-superconductor hybrid structures and shows that certain models can be simulated in a simplified setup. Section~\ref{sec:ExperiProbes} describes how key quantities like the excitation spectrum or dynamic correlation functions can be detected. Section~\ref{sec:Applications} highlights possible applications of the quantum simulator. Finally, Sec.~\ref{sec:discussion} discusses related work and presents an outlook.

\section{Basic Setup}\label{sec:setup}

In this section we outline the basic principles behind our setup.
A more detailed discussion of the experimental realization and parameters is deferred to Sec.~\ref{sec:ExpRealization}.
We start by describing the simplest variant of our proposal: each effective spin-1/2 is represented by a mesoscopic superconducting island hosting four Majorana states.  
We show that arbitrary trivalent lattices can be built out of these sites.
Next, we extend the setup to lattices with larger coordination number, higher effective spins, and non-planar graphs (\emph{i.e.}, lattices with bonds that cross).
Our scheme relies on a perturbative expansion, and so we conclude with a discussion of the higher order corrections not explicitly considered.

\subsection{Spin-1/2 models on trivalent lattices}\label{sec:3bond}

The primary component of our proposal is the mesoscopic superconducting island. 
We take the superconducting gap to greatly exceed all energy scales of the system. 
The next most dominant term in the Hamiltonian is the charging energy:
\eq{\label{eqn:ChargingHam}
H_{C,i}&={e^2\o 2C_i}\left(n_{e,i}-{Q_{0,i}\o e}\right)^2,
}
where $i$ labels the island, $n_{e,i}$ is the number operator of the electrons on the superconductor, and $C_i$ is the capacitance.
The background charge, $Q_{0,i}$, is controlled by a gate voltage, and we assume that it has been tuned to favour an integer number of charges: $Q_{0_i}/e=N_i+\d N_i$, where $N_i\in \mathds{Z}$ and $\abs{\d N_i}\ll1$.
As a result, there is an energy cost of $e^2/2C$ when $n_{e,i}= N_i\pm1$.
We work in the regime where $e^2/2C$ dominates all energy scales save the superconducting gap, and it follows that the low energy physics can be understood by working entirely within the groundstate manifold of $H_{C}$.

When $N_i$ is an even number and the superconductor is topologically trivial, the optimal charge subspace is non-degenerate.
Conversely, topologically non-trivial superconductors may possess additional degrees of freedom in the form of Majorana zero modes, $b_n$.
Their defining properties are
\eq{\label{eqn:MajDef}
\{b_n,b_{n'}\}&=2\d_{nn'},
&
b_n^\dag&=b_n.}
Because of the reality condition, a single Majorana cannot be associated to an occupation number and therefore does not constitute a proper excitation.
Nevertheless, Majoranas always appear in pairs, and so a two-dimensional Hilbert space labelled by the occupation of a complex fermion $f_n=(b_{2n}+ib_{2n-1})/2$ can always be defined.
In the absence of a charge constraint, the topological character of the superconductor supporting these modes guarantees that the two states $\Ket{0}_n$ and $\Ket{1}_n=f_n^\dag\Ket{0}$ are \emph{nearly} degenerate, with an energy splitting that decreases exponentially with the separation of the two modes.
When $N_m$ Majorana zero modes are present in the superconductor, the degeneracy becomes $2^{N_m/2}$.
Finally, fixing the total charge of the system removes one degree of freedom and reduces the dimension of the groundstate manifold to $2^{N_m/2-1}$.

\begin{figure*}[t]
\centering
\includegraphics[width=.95\textwidth]{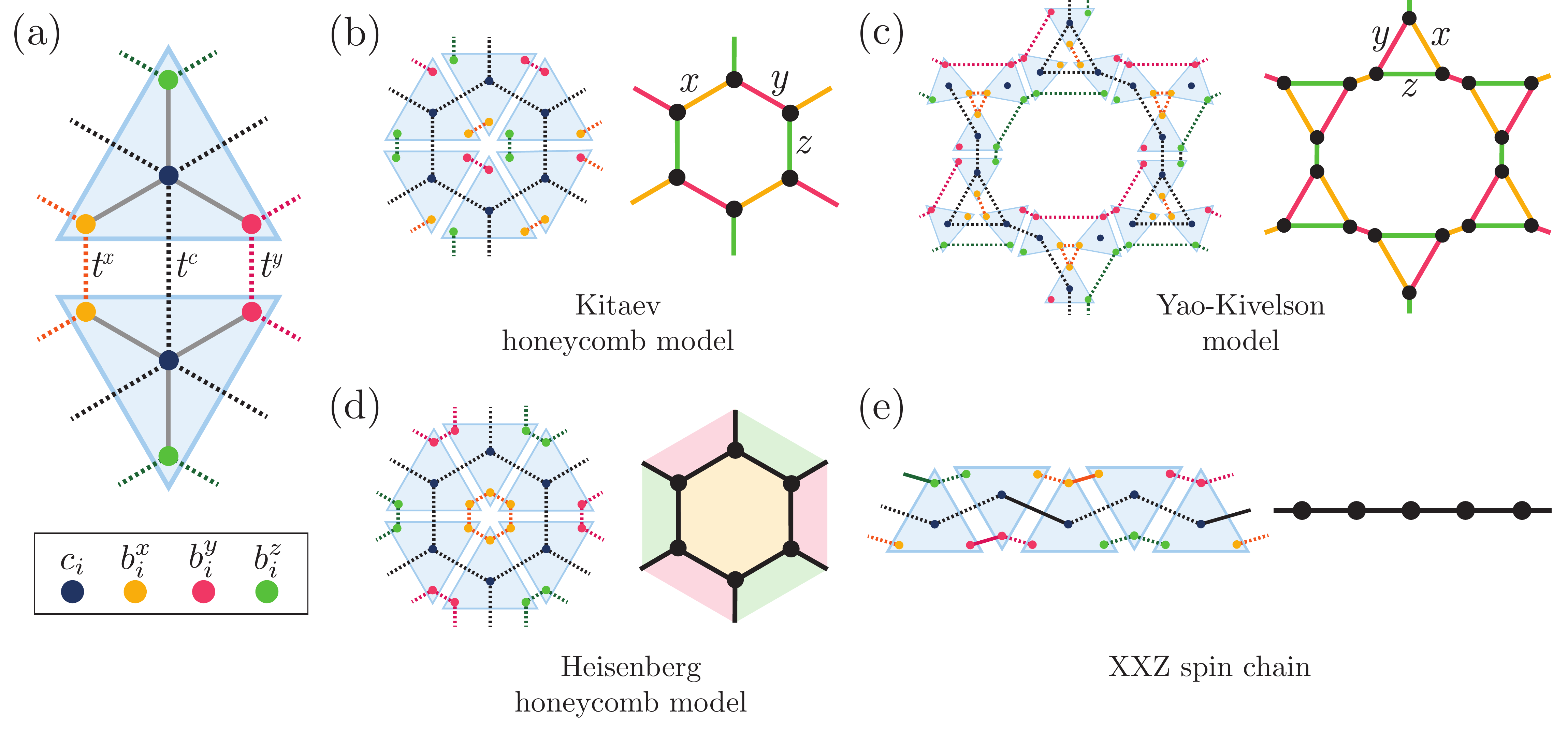}
\caption{Basic spin simulation setup. 
(a) Two Majorana islands interacting through an $\ell_{xy}$-bond. The islands are represented by the blue triangles and the three grey lines depict the proximity-coupled semiconductor wire hosting Majorana zero modes at their ends -- these are omitted in all subsequent diagrams. 
The inset Majoranas shows which coloured circle is associated to which of the four Majoranas: $c$, $b^{x,y,z}$.
The remaining diagrams illustrate Majorana network representations of different spin models:
(b) the Kitaev honeycomb model, (c) the Yao-Kivelson model, (d) the Heisenberg honeycomb model, (e) an XXZ spin chain.
}
\label{fig:trivalentModels}
\end{figure*}

Majoranas may be found both localized at the ends of 1$d$ topological superconductors and localized within vortices in 2$d$ topological superconductors. 
Inspired by recent experimental success in the former system \cite{Albrecht2016}, we propose the setup shown in Fig.~\ref{fig:trivalentModels}(a).
The two blue triangles represent mesoscopic, trivial superconductors, while the gray lines depict topological superconducting wires.
Together, they comprise a single superconducting island with charging energy $H_{C,i}$.
Most importantly, a total of four Majoranas, labelled by $c_i$, $b_i^x$, $b_i^y$, and $b_i^z$, is also present on each island as indicated by the coloured dots.

A specific experimental realization of this device is discussed in Sec.~\ref{sec:ExpRealization}.
Nevertheless, we stress that our protocol is not restricted to those systems or even to ones containing 1$d$ topological superconductors.
The only mandatory component is a superconducting island with a large charging energy and four Majoranas.
For this reason, the wires are omitted in subsequent diagrams.

With four Majoranas and the charge constraint, each island is two-fold degenerate and the parity operator,
\eq{\label{eqn:constraint}
p_i=c_ib_i^xb_i^yb_i^z=\pm1
}
takes a fixed value.
Depending on how the Majoranas are defined, $p_i=+1$ may correspond to either an even or an odd parity.
We discuss this issue later in this section.

In the fixed-parity subspace, it is not difficult to show that the operators
\eq{\label{eqn:PauliDef}
\s^a_i&={i}p_ic_ib_i^a,
&
a=x,y,z.
}
reproduce the spin algebra. 
Further, the identity $p_i=c_ib_i^xb_i^yb_i^z$ implies that the spin operators may equivalently be represented solely in terms of $b$-Majoranas:
\eq{\label{eqn:PauliDef2}
\s^a_i&=-i\ep^{abc}b_i^bb_i^c,
}
where repeated superscripts are summed.
As we discuss below, this manner of representing spin operators was used by Kitaev to exactly determine the groundstate of the honeycomb model \cite{Kitaev2006}.
We employ this representation in a manner converse to Kitaev: instead of representing physical spins with Majoranas, we represent the physical Majoranas as spins.

Interactions between neighbour spins originate from weak tunnelling between the Majoranas on separate islands.
These inter-island bonds can be differentiated by which $b$-Majoranas are interacting.
For instance, the two islands in Fig.~\ref{fig:trivalentModels}(a) are interacting through the $b_i^x$, $b_i^y$, and $c_i$ Majoranas, and so we call this an ``$xy$-bond" or $\ell_{xy}$.
Denoting the tunneling amplitudes by $t_{ij}^x$, $t_{ij}^y$, and $t_{ij}^c$, we write the tunneling Hamiltonian as
\eq{\label{eqn:tunHam}
H^{(xy)}_{\mathrm{tun}\! ,ij}&=
g_{ij}^x i b_i^xb_j^x+g_{ij}^yi b_i^yb_j^y+g_{ij}^cic_ic_j,
\nt
g_{ij}^a&={1\o2}\left[t_{ij}^a e^{i\left( \phi_i-\phi_j\right)/2}+\overline{t_{ij}^a}e^{-i\left(\phi_i-\phi_j\right)/2}\right],
} 
where $a=x,y,c$ and $\phi_i$ and $\phi_j$ are the superconducting phases on either island.
We are interested in the regime where $|t_{ij}^a|\ll e^2/2C$, justifying a perturbative expansion.
Because the problem is dominated by the charging Hamiltonian in Eq.~\eqref{eqn:ChargingHam}, the island  does not have a definite phase.
Instead, $e^{i\phi_i/2}$ is an operator that creates a charge on the island at site $i$. 
It follows that, projecting to the ground state manifold of $H_{C,i}$ and $H_{C,j}$ annihilates $H^{(xy)}_{\mathrm{tun},ij}$, and the leading order contribution appears at second order in perturbation theory:
\eq{
H^{(xy)}_{ij}
&=-{1\o2}\left({1\o E_{C,ij}}+{1\o E_{C,ji}}\right)\Big[ 
\mathrm{Re}\left(t_{ij}^x\overline{t^y_{ij}}\right)b_i^xb_i^yb_j^xb_j^y
\nt
&\quad
+\mathrm{Re}\left(t_{ij}^x\overline{t_{ij}^c}\right)c_ib_i^xc_jb_j^x
+\mathrm{Re}\left(t_{ij}^y\overline{t_{ij}^c}\right)c_ib_i^yc_jb_j^y\Big],
}
where a constant has been omitted.
Here, we have defined $E_{C,ij}=U_{i+}+U_{j-}$, where $U_{i\pm}=H_{C,i}(n_i\pm1)-H_{C,i}(n_i)$ is the energy cost of removing or adding an electron to the $i$th island. 
In what follows, we assume that $H_{C,i}=H_C$ is the same for all islands and write $E_C=U_++U_-$; we frequently refer to $E_C$ as the ``charging energy."
We next use Eqs.~\eqref{eqn:PauliDef} and~\eqref{eqn:PauliDef2} to rewrite the Majorana bilinears in terms of spin operators:
\eq{\label{eqn:HxyEff}
H^{(xy)}_{ij}&=
{1\o4}\sum_{a=x,y,z}J_{ij}^a\s^a_i\s^a_j,
}
where the exchange couplings have been defined as
\eq{\label{eqn:HeisenbergCouplingDef}
\cJ_{ij}^{x,y}&
={4\o E_C}p_ip_j\mathrm{Re}\left(t_{ij}^c\overline{t_{ij}^{x,y}}\right),
&
\cJ_{ij}^z&
={4\o E_C}\mathrm{Re}\left(t_{ij}^x\overline{t_{ij}^y}\right).
}
It is not difficult to see that the analogous expression for interactions across $yz$- and $zx$-bonds are identical to Eq.~\eqref{eqn:HxyEff}, but with appropriately redefined $J_{ij}^{x,y,z}$.
The total Hamiltonian of the network is
\eq{\label{eqn:Heff}
H_{\mathrm{eff}}&=\sum_{\ell_{xy}}H_{ij}^{(xy)}+\sum_{\ell_{yz}}H_{ij}^{(yz)}+\sum_{\ell_{zx}}H_{ij}^{(zx)},
}
where $\ell_{ab}$ with $a,b=x,y,z$ indicates which type of bond is being summed over.

We further note that it is also possible to induce an onsite transverse Zeeman field by coupling two Majoranas within an island:
\eq{\label{eqn:EffZeeman}
H^{(z)}_{Z,i}=ih_Z b_i^x b_i^y=-h_Zp_i \s^z_i.
}
Here, a magnetic field in the $z$-direction is simulated, but the other directions are equally possible.
As opposed to the inter-island interactions considered above, the tunnelling amplitude enters directly into the effective Hamiltonian and is not suppressed by the charging energy.
It is therefore essential that $h_Z$ be extremely small. 

Both the Heisenberg couplings in $H_{\mathrm{eff}}$ and the effective Zeeman fields in $H_Z$ depend sensitively on the phases of the tunnelling amplitudes, making it imperative to understand the relation $H_{ij}^{(ab)}$ and $H_{Z,i}$ have to the physical observables in the system. 
We start by observing that the expressions in Eqs.~\eqref{eqn:tunHam} and~\eqref{eqn:EffZeeman} follow from hopping between the \emph{electrons} of the model.
For example, the interaction between the two $x$-Majoranas in Eq.~\eqref{eqn:tunHam} arises from
\eq{\label{eqn:tunHamEl}
\tilde{H}^{(x)}_{\mathrm{tun}\!,ij}=-{\tilde{t}_{ij}^x}\left[\(f_{i}^x\)^\dag f_{j}^x+h.c.\right],
}
where $\tilde{t}_{ij}^x$ is real, and $f_{i}^x$ and $f_{j}^x$ are electron operators at the $x$-ends of wires on islands $i$ and $j$ respectively.
At low energies relative to the superconducting gap, the Majoranas are the only modes that contribute to tunneling processes. 
Further, since they are localized operators, they have significant overlap only with electron orbitals in their proximity (\emph{e.g.}, the end of the wire).
It follows that the replacement $f_{i}^x\to u_i^xe^{-i\phi_i/2}e^{i\th_i^x/2}b_i^x$ is appropriate in Eq.~\eqref{eqn:tunHamEl}.
Here, $u_i^x$ is a positive number that parametrizes the overlap of the Majorana and the fermion.
As above, $\phi_i$ represents the superconducting phase of the $i$th island, and it follows that $e^{-i\phi/2}$ is the operator that removes a charge from island $i$.
Importantly, an additional phase contribution, $e^{i\th_{i}^x/2}$, is also present. 
Unlike the superconducting phase, this term is simply a number, and its value that depends both on the spin-orbit direction, how the system was prepared, and a gauge choice.
The latter follows from the observation that since changing the sign of a Majorana does not affect its defining relations in Eq.~\eqref{eqn:MajDef}, there is an ambiguity in the definition of $e^{i\th_i^x/2}$.
The choice of sign ultimately determines to whether the positive eigenvalue of the parity operator in Eq.~\eqref{eqn:constraint} corresponds to the even or odd fermion parity sector of the island.
The tunnelling amplitude in Eq.~\eqref{eqn:tunHam} is related to $\tilde{t}^x$ as
\eq{\label{eqn:t-ttilde}
t_{ij}^x=-2{i}\tilde{t}_{ij}^x\,u_i^xu_j^xe^{i(\th_i^x-\th_j^x)/2}.
}
The experimental parameters that determine these phases are discussed in greater detail in Sec.~\ref{sec:ExCouplingSign}.%

Nevertheless, one can verify that if two islands islands have the same parity $p_i=1$, by tuning the phases $\th^a_i$ appropriately, the three exchange couplings, $J^a_{ij}$, can realize any combination of signs.
In particular, this means that both antiferromagnets and ferromagnets can be simulated. 
Unless otherwise noted, we assume that $t^a$, $\th^a_i$, and $\th^a_j$ are such that $J^a_{ij}>0$.

\subsubsection{Examples}\label{sec:HamExamples}

The scheme just described allows for the simulation of a number of different models, some of which we now discuss.
Our primary constraint is the planarity of the interactions: it is not reasonable to propose a setup in which tunneling amplitudes ``cross" since there is no justifiable manner to preclude all four Majoranas involved from interacting with one another, resulting in large interactions between Majoranas within a single island.
This restriction allows us to simulate lattices with at most coordination number three.

The most direct application of this method is the Kitaev honeycomb model \cite{Kitaev2006}:
\eq{\label{eqn:KitaevHam}
H_K&={1\o4}\bigg(\sum_{\ell_x}\cJ_{ij}^x\s^x_i\s^x_j+\sum_{\ell_y}\cJ_{ij}^y\s^y_i\s^y_j+\sum_{\ell_z}\cJ_{ij}^z\s^z_i\s^z_j\bigg).
}
where the $x$-, $y$-, and $z$-bonds are indicated by the coloured bonds of the lattice on the right of Fig.~\ref{fig:trivalentModels}(b).
Using the relations in Eq.~\eqref{eqn:PauliDef}, Kitaev showed that $H_K$ is exactly solvable \cite{Kitaev2006}, and so it should not be surprising that it can be simulated using physical Majoranas.
A potential realization of this system using Majoranas is shown to the left of Fig.~\ref{fig:trivalentModels}(b).
We note that in this case, fixing the signs of the tunnelling amplitudes or the parities of the islands is not necessary as the sign of the exchange couplings $J^a_{ij}$ does not alter the groundstate.

The honeycomb model is just one of a family of exactly solvable models with spin liquid groundstates: any system of spin-1/2's on a trivalent lattice with bonds labelled by $x$, $y$, and $z$ and interacting through $H_K$ is amenable to the same methods used to solve the honeycomb model  \cite{Yao2007,Yang2007,Tikhonov2010,Kells2011}, and our setup is naturally equally well-equipped to simulate these models.
One example is the Yao-Kivelson model where the spins take positions on a decorated honeycomb lattice, as shown in Fig.~\ref{fig:trivalentModels}(c) on the right \cite{Yao2007}.
For an appropriate choice of coupling, this model realizes a chiral spin liquid.
For instance, the groundstate spontaneously breaks time-reversal symmetry and supports non-Abelian excitations in the form of non-trivial fluxes.
While the appearance of non-Abelian statistics in a system whose fundamental components are Majoranas may not be surprising, the non-Abelian nature of the flux excitation is an emergent phenomenon, independent from the basic constituents of the spins being simulated.
(This can be compared with the distinction between the fermionic spinon in a $\Zt$ spin liquid and the electron of the ultraviolet theory.)

Since the Kitaev models interact exclusively through Ising interactions, it is clear that the Majorana network we propose can also realize Ising models on trivalent lattices as well as chains.
 
While the simulation of exactly-solvable and Ising models is arguably not particularly interesting, our scheme easily generalizes to unsolvable models.
In particular, the familiar Heisenberg Hamiltonian is obtained when all neighbouring Majoranas couple with the same tunneling strength:
\eq{
H_H&={\cJ\o4}\sum_{\Braket{ij}}\v{\s}_i\cdot\v{\s}_j.
}
In Fig.~\ref{fig:trivalentModels}(c) the Heisenberg model on the honeycomb lattice is shown.
Unlike the Kitaev models, each bond of the Heisenberg model requires three Majoranas to interact.
As a result, the internal orientation of the Majoranas within each island imposes a further restriction on the lattices we can simulate: not only must they be (at most) trivalent, but they must also be \emph{bipartite}. 
As a result, at this point, the scheme we have outlined is not able to simulate the Heisenberg model on the decorated honeycomb lattice shown in Fig.~\ref{fig:trivalentModels}(c).

A crucial feature of our setup is that the SU(2) symmetry of $H_H$ is only achieved through the fine-tuning of the values of the inter-site couplings.
We discuss how this may be implemented experimentally in Sec.~\ref{sec:SU2sym}.
The absence of an exact SU(2) symmetry actually has some positive consequences.
Notably, it allows us to simulate XXZ-type models,
\eq{\label{eqn:XXZHam}
H_\mathrm{XXZ}&={\cJ\o4}\sum_{\Braket{ij}} \Big( \s^x_i\s^x_j + \s^y_i\s^y_j + \gamma \s^z_i\s^z_j \Big),
}
where $\g\neq0$.
This can be accomplished by the following choice of couplings: on both $yz$- and $zx$-bonds, let the magnitudes of $t^y$ and $t^x$ differ from the magnitudes of $t^c$ and $t^z$, \emph{i.e.} $(t^a,t^z,t^c)=(t/\g,t,t)$, $a=x,y$.
In contrast, let $t^c$ differ from the other couplings on $xy$-bonds instead: $(t^x,t^y,t^c)=(t,t, t/\g)$.
This is illustrated for a spin chain in Fig.~\ref{fig:trivalentModels}(e).
There, the anisotropic couplings (those with coupling strength $t/\g$) are indicated by a straight line, while those with strength $t$ are indicated with dashed lines

More interesting yet, is the our ability to simulate the Heisenberg-Kitaev model \cite{KHmodel_Chaloupka10}:
\eq{\label{eqn:HKham}
H_\mathit{HK}&={\cJ_H \o4}\sum_{\Braket{ij}}\v{\s}_i\cdot\v{\s}_j 
+
{\cJ_K\o4}\sum_{\ell_a}\s^a_i\s^a_j.
}
This can be obtained by the natural generalization of the recipe above.
This model is most interesting when $\cJ_H>0$ and $\cJ_K<0$, and so the signs of the Kitaev exchange couplings do matter in this case.

\subsection{Four-bond vertex}\label{sec:4bond}

\begin{figure}
    \centering
    \includegraphics[width=0.47\textwidth]{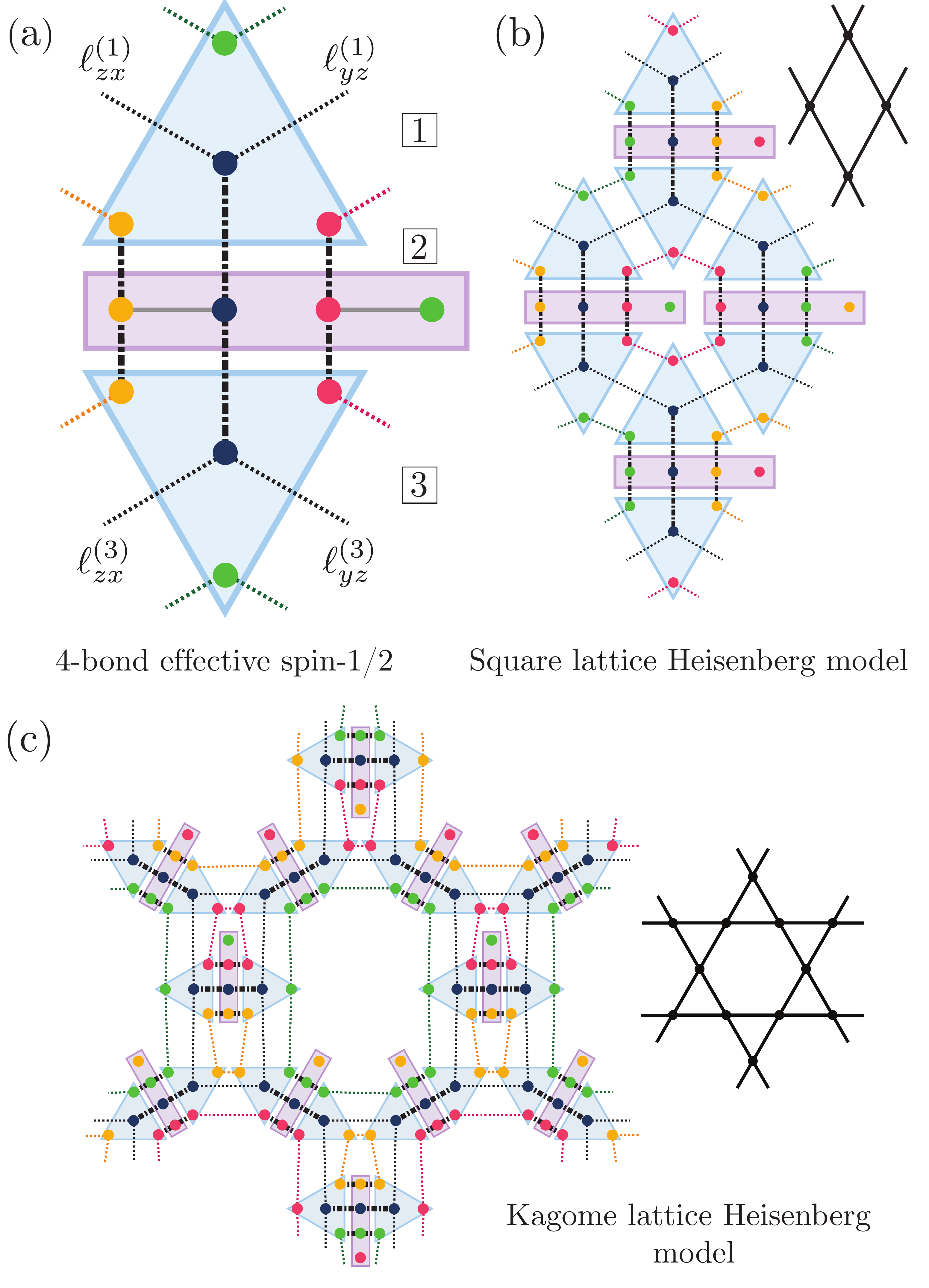}
    \caption{
    (a) The three-island vertex proposed in Sec.~\ref{sec:4bond}. As opposed to the single-island vertex, it can couple to four bonds neighbours. Here, these bonds are indicated by the labels $\ell_{zx}^{(1,3)},\,\ell_{yz}^{(1,3)}$.
    The thick, black dash-dotted line represent the strong tunneling strength within the vertex, $t_\mathrm{vert}$, while the smaller, dotted lines represent tunneling between neighbour vertices with strength $t_\mathrm{bond}$.
    The other two diagrams are Majorana representations of Heisenberg models on (b) the square and (c) the kagome lattices.
    }
    \label{fig:4bond}
\end{figure}

The scheme outlined above is limited to lattices with coordination number three or less and is therefore not applicable to many of the models of greatest interest, such as the kagome Heisenberg model.
Using the simple fact that the groundstate of a (short) Heisenberg chain with an odd number of spins is two-fold degenerate and has total spin 1/2, we extend our setup to accommodate Heisenberg models on four-bond lattices.
(The configuration discussed in this section is actually capable of simulating bipartite lattices with up to coordination number five, but we do not explicitly discuss an example of this form.)

In this scheme, each vertex of the lattice is composed of three Majorana islands arranged in a row $1-2-3$, as shown in Fig.~\ref{fig:4bond}(a).
Islands $1$ and $3$ are each coupled to two other vertices through the bonds labelled $\ell^{(i)}_{zx},\ell^{(i)}_{yz}$, $i=1,3$.
In order to view these three islands as a unit, we assume that the intra-vertex coupling strength is very strong relative to the inter-vertex coupling strength: $t_\mathrm{vert}\gg t_\mathrm{bond}$.

There are several ways to understand why this structure represents a spin-1/2.
In terms of the Majoranas, we observe that the interactions result in the hybridization of the $c$, $b^x$, and $b^y$ Majoranas of each island, allowing the identification $c_{1}\sim c_2\sim c_3$ and so on.
When the ever-present charge constraint is taken into account, we further obtain the non-local identification $b_1^z\sim b_2^z\sim b_3^z$. 
It follows that for energies much lower than $t_\mathrm{vert}$, the three-island vertex should possess the same degrees of freedom as a single spin-1/2 degree of freedom and can be treated in the same way as the single island in the previous section.

This can be shown exactly in limit where both inter- and intra-vertex tunneling amplitudes are much smaller than the charging energy, $E_C\gg t_\mathrm{vert} \gg t_{\mathrm{bond}}$.
To start, we note that since $E_C$ is the dominant energy scale, the arguments of previous section establish that to leading order in $1/E_C$ the behaviour of the charge-neutral sector is controlled by the effective Hamiltonian in Eq.~\eqref{eqn:Heff}. 
It is convenient to rewrite the Hamiltonian as $H_\mathrm{eff}=\sum_i \left( H_{v,i}+H_{b,i}\right)$, where $i$ sums over the vertices and 
\eq{
H_{v,i}&=\cJ_v\left(\v{S}_{i,1}+\v{S}_{i,3}\right)\cdot \v{S}_{i,2},
\\
H_{b,i}&={\cJ_b\o2}\[\v{S}_{i,1}\cdot\left(\v{S}_{j_1}+\v{S}_{j_2}\right)+\S_{i,3}\cdot\left(\S_{j_3}+\S_{j_4}\right)\].\notag
}
Here $\cJ_v\sim t_\mathrm{vert}^2/E_C$, $\cJ_b\sim t_\mathrm{bond}^2/E_C$, and $\S_{i,\xi}=\v{\s}_{i,\xi}/2$, $\xi=1,2,3$ are the spin operators corresponding to the Pauli matrices defined in Eq.~\eqref{eqn:PauliDef}. 
The generically indexed spins $\S_{j_1},\S_{j_2},\S_{j_3},$ and $\S_{j_4}$ represent the effective spins on the four nearest-neighbouring islands.

Next, the inequality $t_\mathrm{vert}\gg t_\mathrm{bond}$ implies that $\cJ_v\gg\cJ_b$ as well.
Our final effective Hamiltonian is obtained by projecting onto the groundstate manifold of $\sum_iH_{v,i}$, and this can be done by focusing on a single three-island vertex at a time.
For notational clarity, we suppress the vertex index~$i$.

The groundstate of $H_v$ is determined  by rewriting the spin operators as $\v{S}_{13}=\S_1+\S_3$, $\v{S}_\mathrm{tot}=\S_{13}+\S_2$.
This gives
\eq{
H_v&
={\cJ_v\o2}\left(\v{S}_\mathrm{tot}^2-\S_{13}^2-{3\o4}\right)\cdot
}
It's clear that $H_v$ can be simultaneously diagonalized with respect to the total spins $s_\mathrm{tot}$ and $s_{13}$.
The resulting groundstate manifold, $\mathscr{H}_0$, is two-fold degenerate with quantum numbers $(s_\mathrm{tot},s_{13})=(1/2,1)$.
The energy gap to the next excited state is $\cJ_v$, which is by assumption larger than the bond exchange coupling as required. 

We project onto the groundstate manifold $\mathscr{H}_0$ through the projection operator $\P$.
Letting $\vcS=\P\S_\mathrm{tot}\P$ be the spin operator on $\mathscr{H}_0$, we find
\eq{\label{eqn:projSpin}
\P \S_{1,3}\P&={2\o3}\vcS,
&
\P\S_2\P&=-{1\o3}\vcS.
}
That is, spins~1 and~3 act identically on $\mathscr{H}_0$. 
Performing the same analysis on every site, the effective Hamiltonian is
\eq{\label{eqn:4bondHeff}
H_\mathrm{eff}&=\cJ_\mathrm{eff}\sum_{\Braket{ij}}\vcS_i\cdot\vcS_j,
&
\cJ_\mathrm{eff}&={4\o9}\cJ_b.
}
Higher order terms are suppressed by $\cJ_b/\cJ_v$.

In Figs.~\ref{fig:4bond}(b) and~(c) we illustrate possible realizations of both the square and kagome lattice Heiseberg models.
As with the single-island vertices of the previous section, XXZ-type anisotropies may also be induced for greater frustration.

This proposal may appear experimentally unfeasible since it is seemingly founded on the double limit $E_C\gg t_\mathrm{vert} \gg t_\mathrm{bond}$.
However, we now demonstrate that these constraints can be considerably relaxed by numerically solving the Hamiltonian
\eq{\label{eqn:ExactVertexHam}
H_\mathrm{fluc}&=
\sum_{i=1}^3 H_{C,i} + H_{\mathrm{tun},12}^{(xy)} + H_{\mathrm{tun},23}^{(xy)}+ H_J,
}
where $H_{C,i}$ and $H_{\mathrm{tun},ij}^{(xy)}$ are the charging and tunneling Hamiltonians defined in Eqs.~\eqref{eqn:ChargingHam} and~\eqref{eqn:tunHam} respectively.
We assume that the charging energies, $E_C$, and tunneling amplitudes, $t_\mathrm{vert}$, are the same for all islands and bonds.
$H_J$ is the Josephson Hamiltonian:
\eq{
H_J&=-E_J\Big[
\cos(\phi_1-\phi_2)+\cos(\phi_2-\phi_3)
\Big].
}
So far, this term has not been considered as its contribution is negligible when $E_C$ is very large. However, in the limit of weak tunneling, the Josephson energy is proportional to the square of the tunneling amplitude and so should be taken into consideration once $t_\mathrm{vert}$ ceases to be small.
We write simply $E_J=\ep_Jt_\mathrm{vert}^2$ and solve for different choices of proportionality constant $\ep_J$.
More details are provided in Appendix~\ref{app:4bondSims}.

\begin{figure}
\centering
\includegraphics[width=0.48\textwidth]{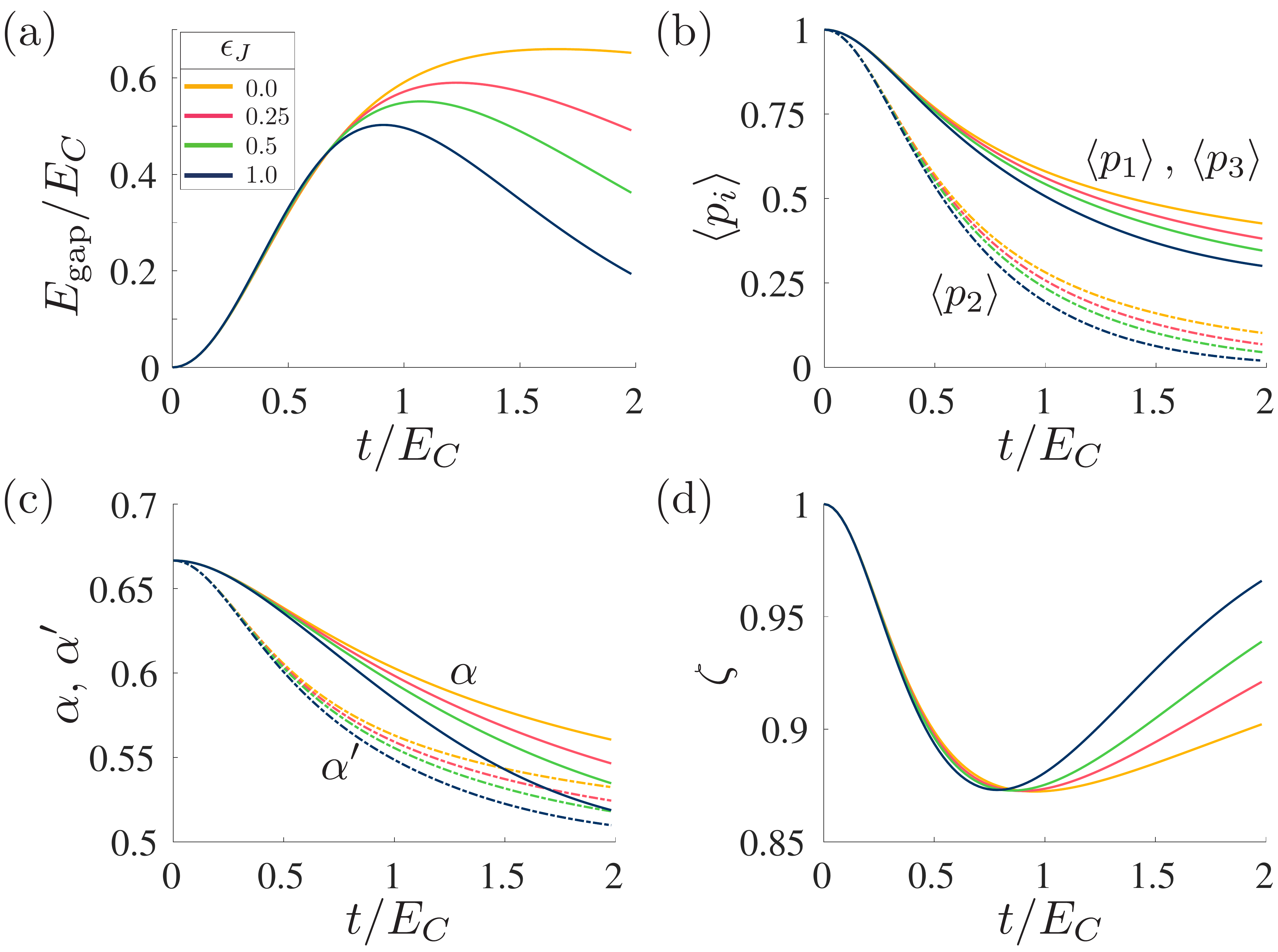}
\caption{Numerical results of three-island vertex simulation. 
Together they demonstrate that the optimal coupling strength is $t_\mathrm{vert}\sim E_C$.
The functions in (a)-(d) are all plotted against the tunneling amplitude $t_\mathrm{vert}$.
The colours are associated with the values of $\ep_J$ provided in the inset of (a).
The functions plotted are: (a) the energy gap between the two-fold degenerate manifold of interest and the excited states; (b) the expectation value of the parity operator on islands~1-3; (c) the proportionality constants $\a$ and $\a'$ of the now-inequivalent spin representations in Eq.~\eqref{eqn:AnisoSpinProj}; (d) the ratio $\zeta=\a'/\a$ parametrizing the anisotropy between spin representations.
}
\label{fig:3IslandVertNumerics}
\end{figure}
Our results are shown in Fig.~\ref{fig:3IslandVertNumerics}.
In (a), we plot the energy gap as a function of the tunneling amplitude.
We find that it reaches a maximum when $t_\mathrm{vert}\cong E_C$.
This demonstrates that the double limit $E_C\gg t_\mathrm{vert}\gg t_\mathrm{bond}$ is not in fact necessary for the three-island vertex to realize a single effective spin-1/2.
The instead, the optimal situation occurs when $E_C\sim t_\mathrm{vert}\gg t_\mathrm{bond}$.

After this point, the Josephson term begins to dominate. 
For large enough $t_\mathrm{vert}$, we find that the excited states to approach the groundstate exponentially as expected.
Physically, the penalty for increasing $t_\mathrm{vert}$ is that inter-island charge fluctuations are allowed. 
This is clearly demonstrated in Fig.~\ref{fig:3IslandVertNumerics}(b), where the expectation value of the parity operator [see Eq.~\eqref{eqn:constraint}] is plotted with respect to $t_\mathrm{vert}/E_C$.
As expected, for all three islands, the parity decreases quite rapidly as $t_\mathrm{vert}$ is increased. 
Since $H_\mathrm{fluc}$ is symmetric under the exchange of islands~1 and~3, their parities are equal, $\Braket{p_1}=\Braket{p_3}$. 
The fact that these expectation values are larger than $\Braket{p_2}$ is in accordance with the reasoning that  because island 2 interacts with two islands while islands~1 and~3 interact with a single island, the parity of island~2 should fluctuate more than the parities of islands~1 and~3 .

The primary consequence of these charge fluctuations is that the two spin representations in Eqs.~\eqref{eqn:PauliDef} and~\eqref{eqn:PauliDef2} cease to be equivalent.
Importantly, a pseudospin operator, $\vcS$, can still be defined because the topological properties of the Majoranas continue to guarantee the twofold degeneracy of the groundstate manifold of the three-island system.
Noting that $H_\mathrm{fluc}$ is symmetric under not only the exchange of islands 1 and 3, but also under permutations of $c$, $b^x$, and $b^y$, we expect
\eq{\label{eqn:AnisoSpinProj}
\a\vcS&=\frac{1}{2}\P\left(-ib^y_jb^z_j,-ib^z_jb^x_j,ic_jb^z_j\right)\P,
\nt
\a'\vcS&=\frac{1}{2}\P\left(ic_jb_j^x,ic_jb_j^y,-ib_j^xb_j^y\right)\P,
}
where $j=1,3$ and $\a$ and $\a'$ are real.
As above, $\P$ represents the groundstate projector.
Since $c_{1,3}$, $b_{1,3}^x$, and $b_{1,3}^y$ interact with Majoranas on island~2, we further infer that operators composed only of these Majoranas should have a smaller overlap with $\vcS$ compared to those that contain $b_j^z$ in their definition, implying that $\a'<\a$.
This is verified in Fig.~\ref{fig:3IslandVertNumerics}(c), where both $\a$ and $\a'$ are plotted against the tunneling amplitude. 
We observe that when $t_\mathrm{vert}$ is zero, $\a=\a'=2/3$, confirming that Eq.~\eqref{eqn:projSpin} holds in the appropriate limit.

One effect of Eq.~\eqref{eqn:AnisoSpinProj} is to renormalize the effective Heisenberg coupling of the inter-vertex Hamiltonian.
While it clearly reduces the interaction strength relative to the value obtained in the $t_\mathrm{vert}/E_c\to0$ limit, this effect is small. 
Even for the largest $t_\mathrm{vert}$ and $\ep_J$ considered, the resulting coupling is still over half the effective coupling in Eq.~\eqref{eqn:4bondHeff}.

A feature of Eq.~\eqref{eqn:AnisoSpinProj} that may be more troubling is that it introduces an inherent anisotropy.
Even supposing that all inter-site tunneling amplitudes have been tuned to have the same magnitude, an SU(2)-symmetric effective Hamiltonian will not be obtained. 
For instance, the exchange couplings corresponding to the $\ell_{yz}^{(1)}$ bond in Fig.~\ref{fig:4bond}(a) will instead be $\cJ_\mathrm{eff}^x= 4\a^2t_\mathrm{bond}^2/E_C$ and $\cJ_\mathrm{eff}^{y,z}=4\a'^2 t_\mathrm{bond}^2/E_C$ in this case (we've assumed that the proportionality constants $\a$, $\a'$ are the same for both three-island vertices).
We parametrize this anisotropy by $\zeta=\a'/\a\leq1$, and this is plotted in Fig.~\ref{fig:3IslandVertNumerics}(d).
It is not too small and can readily be compensated for by an appropriate tuning of the inter-site tunneling amplitudes.

\subsection{Larger coordination number and effective spin}\label{sec:MoreBonds}

By considering vertices composed of more than three islands, the scheme outlined above can be extended not only to vertices with more bonds, but to vertices with larger effective spins as well.
In this section, we work entirely in the regime where each island can independently treated as an effective spin-1/2; as above, this assumes $E_C\gg t_\mathrm{vert}\gg t_\mathrm{bond}$.
We further simplify the problem by assuming a single, uniform intra-vertex exchange coupling, $J_v$.
The vertex is then described by the Hamiltonian
\eq{\label{eqn:HChainVertex}
H_v&=J_v\sum_{\Braket{n,m}}\S_n\cdot\S_m.
}
We let $N_s$ denote the total number of islands in the vertex.

In this section, it is useful to consider both antiferromagnetic and ferromagnetic exchange couplings. 
For the antiferromagnetic case, $J_v>0$, we further assume that the islands are arranged in a bipartite fashion.
We denote the sublattices by $A$ and $B$, and take the number of $A$-sublattice sites to be greater than or equal to the number of $B$-sublattice sites: $N_A\geq N_B$.

The vertex has been reduced to a simple spin problem, and many known results can be applied \cite{auerbach,Marshall55,Lieb62}.
In particular, it can be shown that the groundstate manifold of $H_v$ has total spin
\eq{\label{eqn:GndStateJtot}
s_\mathrm{tot}=
\begin{cases}
{1\o2}\abs{N_A-N_B}, & J_v>0, \\
{1\o2}N_s,  &   J_v<0
\end{cases}
}
and that it is $s_\mathrm{tot}(s_\mathrm{tot}+1)$-fold degenerate, where $s_\ttot$ is the eigenvalue of the total spin $\S_\ttot=\sum_n \v{S}_n$.
Moreover, the groundstate-projected single-site spin operator is proportional to the total spin in the groundstate.
For the ferromagnetic case, the relationship is simply
\eq{
\P \S_n \P = {1\o N_s}\vcS,
}
where we continue denote the operator which projects onto the groundstate manifold by $\P$ and the total spin in the groundstate by $\mathcalbb{S}=\P\S_\mathrm{tot}\P$.
For antiferromagnetic $J_v>0$, the sign of the proportionality constant alternates between sublattices:
 \eq{\label{eqn:ProjSpinSign}
\P\S_n\P&=c_n\vcS,
&
\mathrm{sgn}(c_n)&=\begin{cases}
+1,\quad n\in A
\\
-1,\quad n\in B
\end{cases}.
}

This scheme is limited by the size of the energy gap.
For 1$d$ and quasi-1$d$ chains, the number of sites, $N_s$, increases, ferromagnetic and antiferromagnetic gaps approach zero as $1/N_s^2$ and $1/N_s$, respectively.
When the islands are arranged as a Cayley tree, different gap scalings can be obtained \cite{Changlani13}.

\subsubsection{Larger coordination number}

It is simple to verify that Eqs.~\eqref{eqn:GndStateJtot} and~\eqref{eqn:ProjSpinSign} are consistent with the four-bond vertex presented above.
In Fig.~\ref{fig:ExtSpins}(a), a row of five strongly-coupled islands is shown, and this will behave like an effective spin-1/2 with seven bonds. 
There, islands with a positive proportionality constant ($A$-sublattice) are coloured in blue, while those with a negative proportionality constant ($B$-sublattice) are shown in purple.
This type of vertex can be employed to realize the triangular lattice Heisenberg model.

When attempting to increase the coordination number, it is the natural to choose a configuration which minimizes the number of sites $N_s$. 
Different connectivity choices may result in different gaps as well.
For instance, as we saw, the three-island vertex has an energy gap $\cJ_v$. 
This decreases to $E_{\mathrm{gap}}\cong 0.72\cJ_v$ for a five-spin chain and $E_\mathrm{gap}\cong 0.56\cJ_v$ for a seven-spin chain.

\subsubsection{Larger effective spin}

\begin{figure}
\centering
\includegraphics[scale=.15]{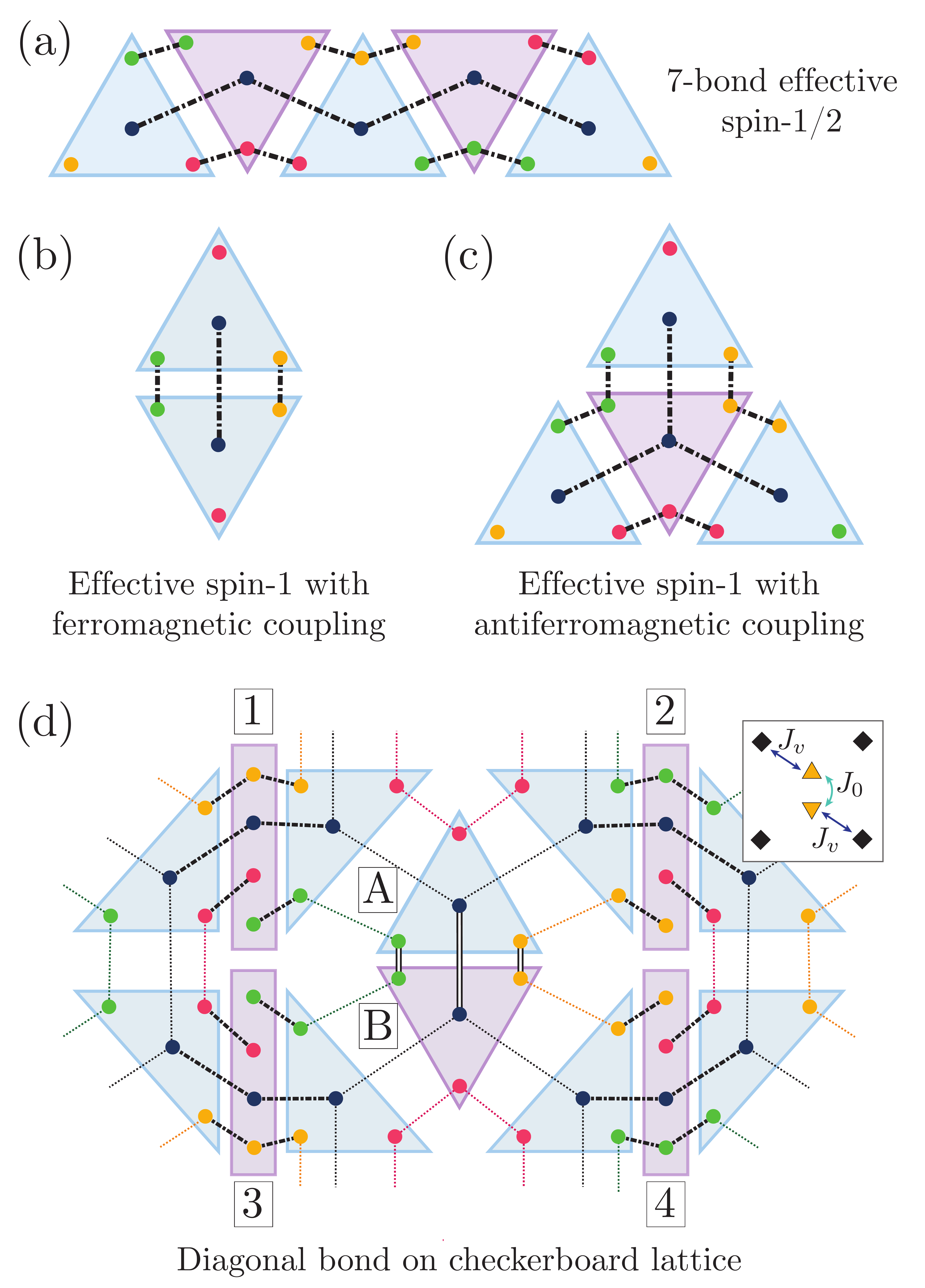}
\caption{Further applications of spin addition. 
(a) A seven bond effective spin-1/2. (b) and (c) both depict effective spin-1 vertices. The couplings in (b) are ferromagnetic, while those in (c) are anti-ferromagnetic. 
(d) An illustration of a portion of a checkerboard lattice model. 
The two islands labelled A and B in the centre are strongly coupled through an antiferromagnetic exchange term so that their groundstate is a singlet.
Through second order perturbation theory they mediate interactions between islands 1 and~4, islands~2 and~3, islands~1 and 2, and islands~3 and~4. 
A simplified schematic is shown in the inset.
}
\label{fig:ExtSpins}
\end{figure}

There are multiple ways to obtain a given effective spin, and the optimal choice depends on whether it is easier to engineer exchange interactions that are ferromagnetic, antiferromagnetic, or a mixture of the two [see Sec.~\ref{sec:ExCouplingSign} below for more discussion]. 
For instance, both the two-island vertex depicted in Fig.~\ref{fig:ExtSpins}(b) and the 4-island vertex in Fig.~\ref{fig:ExtSpins}(c) simulate effective spin-1 degrees of freedom. 
The former corresponds to having ferromagnetic exchange couplings and the energy barrier to the excited states is $\abs{J^\mathrm{FM}_v}$, while the latter configuration requires antiferromagnetic exchange couplings and results in an energy gap $J^\mathrm{AF}_v/2$.

\subsection{Non-planar bonds}\label{sec:DiagBonds}

We have emphasized the requirement that no bonds between Majoranas cross.
This would appear to prohibit the realization of any two-dimensional model with next-nearest neighbour couplings.
However, in this section we demonstrate that the simple spin manipulations exploited in the previous section present one way of overcoming this obstacle.

Consider the setup in Fig.~\ref{fig:ExtSpins}(d). 
There, four 3-island vertices (labelled 1-4) are  arranged in a square and interacting with both their nearest and next-nearest neighbour bonds through the mediation of a two-island singlet (composed of two islands labelled A and B).
The inset provides a simplified schematic of the image.
To see how this works, we consider the regime where the four-bond vertices  and the two center islands can each be treated as a single effective spin.
The interactions between islands~1, 4, A, and B are described by $H_\mathrm{diag}=H_0+H_1$ where
\eq{
H_0&=J_0 \S_A\cdot\S_B, 
\nt
H_1&=J_b\( \S_A\cdot\S_1+\S_B\cdot\S_4\).
}
We assume that $J_0\gg J_v>0$ and project onto the groundstate manifold of $H_0$.
To second order in perturbation theory, the effective Hamiltonian is
\eq{
H_\mathrm{eff}&=J_\mathrm{diag}\S_1\cdot\S_4,
&
J_\mathrm{diag}&={J_b^2\o 2J_0}.
}
This demonstrates our claim that such singlet-islands can help induce diagonal bonds.

There are some caveats. 
First, the coupling induced by hopping between islands 1 and~2 is ferromagnetic in this limit. 
This can sometimes be avoided by changing the relative orientations of the islands, but embedding it into a full lattice can be complicated.
Further, the three-island vertices do not possess enough free bonds to simulate a full square lattice with next-nearest neighbour bonds. 
The best that can be simulated is a checkerboard square lattice with next-nearest neighbours across every other plaquette; for a realization of the $J_1-J_2$ Heisenberg model on the square lattice, vertices composed of more than three islands are needed.

\subsection{Higher order corrections}\label{sec:higher-order}

So far, we have demonstrated that the effective Hamiltonian of the Majorana island network we propose reproduces a variety of Heisenberg Hamiltonians at leading order.
While the derivation of Sec.~\ref{sec:3bond} indicates that all corrections to $H_\mathrm{eff}$ are suppressed by powers of $t/E_C$ (where $t$ is the typical tunneling strength between islands and $E_C$ the charging energy),
one may nevertheless worry that these terms are  physically relevant.

One concern is that even after a model has been fine-tuned to return an SU(2)-symmetric Heisenberg Hamiltonian at leading order, terms depending on the internal orientation of the Majoranas may be generated at higher orders, potentially breaking both SU(2) and spatial symmetries. 
For example, in the Heisenberg model on the honeycomb lattice, one of three colours can be associated to each hexagon.
As an example, the hexagon shown in Fig.~\ref{fig:trivalentModels}(d) can be associated with orange, as shown on the right of diagram.
These colours correspond to different directions in spin space and imply an anisotropy that should ultimately be manifest in the effective Hamiltonian.  

We show that these symmetry-breaking contributions can be neglected up to $\mathcal{O}\left(t^{L-1}/ E_C^{L-2}\right)$, where $L$ is the number of bonds in the smallest loop in the lattice (\emph{e.g.}, $L=6$ for the honeycomb lattice).

We begin by noting that higher-order contributions to the effective Hamiltonian are generated in the same manner we used to obtain $H_\mathrm{eff}$ in Eq.~\eqref{eqn:Heff}.
There, $H_\mathrm{eff}$ was obtained through a two-step hopping process: a Majorana from island $i$ tunnels to a neighbouring island $j$, altering the charge on both islands $i$ and $j$, and taking them out of the groundstate manifold.
To restore the charge, a Majorana must subsequently hop from island $j$ to island $i$.
As we saw, this ultimately results in a contribution of the form $J^a_{ij}\sigma^a_i \sigma^a_j$ for $a=x,\,y,$ or $z$.
Generalizing to higher orders, it's evident that to satisfy the charge constraint, all hopping processes containing fewer than $L$ steps necessarily backtrack on themselves.
We conclude that the $n$th order contribution to the effective Hamiltonian is given by a sum of terms of the form
\eq{\label{eqn:nthOrderCorrection}
H^{(n)}_{\mathrm{corr}}&\sim E_C\left( t\o E_C\right)^n \prod_{\Braket{ij}\in \mathcal{C}_n} \left(\sum_{a=x,y,z} \s^a_i\s^a_j \right)
}
where $n$ is an even number smaller than $L$ and $\mathcal{C}_n$ is half of an $n$-step path that backtracks on itself. 
Save for models containing triangular loops, this demonstrates that the leading correction to $H_\mathrm{eff}$ is of order $t^2/E_C^2$.

When the network includes multi-island vertices or singlet-mediated hopping, $t^n$ should be replaced with the appropriate powers of $t_\mathrm{bond}$, $t_\mathrm{vert}$, and $t_\mathrm{sing}$.
An important observation is that since none of the multi-island vertices we construct in this paper contain internal loops, all higher-order SU(2) symmetry-breaking terms are generated solely via virtual hopping around a true loop of the lattice being simulated.
This means that, in the absence of inter-vertex hopping, the relations of Secs.~\ref{sec:4bond} and~\ref{sec:MoreBonds} are not modified.

\section{Experimental Realization}\label{sec:ExpRealization}

Considerable experimental efforts to build platforms for topological superconductor networks based on superconducting hybrid systems are currently underway. Significant progress has been made for various different platforms, including semiconductor nanowires \cite{Mourik2012,Albrecht2016}, two dimensional electron systems \cite{Shabani2016,Nichele2017}, and atomic chains \cite{Nadj-Perge2014,Ruby2015}, as well as quantum Hall \cite{Lee2017}, quantum spin Hall \cite{Hart2014}, and quantum anomalous Hall \cite{He2017} systems.

Semiconductor-based platforms have shown compelling evidence for the presence of Majorana states and their exponential edge localization in tunneling spectroscopy \cite{Mourik2012,Zhang2018,Albrecht2016,Nichele2017}. Moreover, InAs wires with epitaxial Al shells exhibit clear signatures of topological superconductivity in the Coulomb blockade regime \cite{Albrecht2017}.

Based on these encouraging results, proposals have been made to construct a full-scale quantum computer using Majorana islands, ranging from full architectures based on surface codes \cite{Vijay2015,Landau2016} or color codes \cite{Litinski2017} to concrete experimental geometries to implement qubit operations \cite{Karzig2017,Plugge2017,Vijay2016}. Nevertheless, building a universal quantum computer remains challenging as it requires a large number of qubits to implement various logical gates as well as error correction protocols. In Ref.~\citenum{Litinski2017}, the overhead for a fully operational quantum computer was estimated to be 500 physical qubits per logical qubit. 

Motivated by recent experimental progress, we focus on proximity-coupled semiconducting wires as a platform for the quantum simulator. Such a platform can be realized either by assembling epitaxially-grown nanowires with superconducting shells \cite{Krogstrup2015} or by defining superconducting wires in a two-dimensional electron gas by lithography and gating \cite{Nichele2017,OFarrell2018}. The semiconductor platforms have the additional advantage that other building blocks, such as quantum dots serving as control knobs or external probes, can be added with little cost.

\subsection{Basic experimental setup}

\begin{figure}[t]
 \includegraphics[width=.47\textwidth]{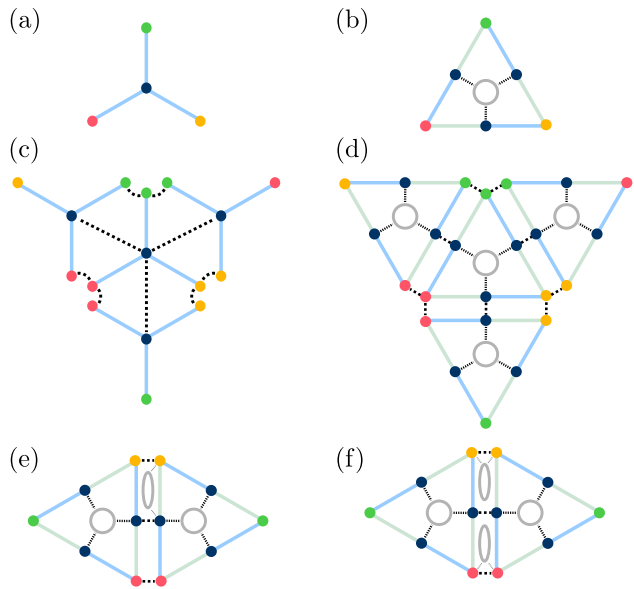}
\caption{(a), (b) Proposed experimental setup for a Majorana island consisting of proximity coupled semiconducting wires hosting Majorana states (colored circles). Light blue (light green) segments correspond to the topological (trivial) phase. Gray circles represent quantum dots and black dashed lines indicate tunnel couplings. (c), (d) Effective Heisenberg coupling realized by tunnel couplings between neighboring islands. (e), (f) Additional quantum dots can be used for measurements or the implementation of Zeeman fields as detailed in Sec.~\ref{sec:ExperiProbes}.}
\label{fig:exp_setup}
\end{figure}

We envision a design for the Majorana island, the elementary building block of the simulator,  as depicted in Fig.~\ref{fig:exp_setup}(a). The island comprises three topological superconductor wires which are connected at one end so that the three Majoranas hybridize to a single Majorana and a fermion with a sufficiently large gap. Moreover, the superconductors should be well connected, such that the three wires have a mutual charging energy and the individual charging energies are suppressed.
Fig.~\ref{fig:exp_setup}(c) shows how the island is connected to neighboring islands realizing a three-bond vertex of Heisenberg couplings. Majoranas of the same type are connected by tunnel couplings. The geometry allows the outer Majoranas $b^{x,y,z}_i$ to be placed nearby the corresponding Majoranas of neighboring islands, and the tunnel couplings can simply be realized with quantum point contacts. Coupling the central $c$-Majoranas, however, requires tunnel couplings across distances comparable to the length of the topological segment. 

Tunnel coupling across longer distances could be realized, e.g., by elongated quantum dots that are tuned off resonance. The spatial extension of gate-defined quantum dots is constrained by the fact that the gap of the dot should exceed the charge gap in the superconducting islands. 
Alternatively, when the separation is too large to be bridged by gate-defined quantum dots, tunnel couplings can be realized by a single topological superconducting wire with a sizable charging energy. The two end Majoranas form a nonlocal quantum dot, which can have a charge gap exceeding the pairing strength even for micron-sized wires.

An alternative setup is depicted in Fig.~\ref{fig:exp_setup}(b). In this case, the island comprises three wires with six Majoranas altogether. Again, all wires are connected to ensure a mutual charging energy. In addition, a central quantum dot tuned off-resonance ensures that the three central $c$-Majoranas are strongly coupled. This results in a single Majorana state delocalized over the three sites. The mutual coupling strengths of the $c$-Majoranas should ideally exceed the gap set by the charging energy. While this setup requires an extra quantum dot, it is more symmetric and all tunnel couplings connecting different islands can be fabricated in the same way [see Fig.~\ref{fig:exp_setup}(d)].

While this basic architecture is sufficient to realize the spin models discussed in Sec.~\ref{sec:setup}, additional elements can be added as probes or control knobs. In the simplest case,  quantum dots can be placed between neighboring islands as shown in Figs.~\ref{fig:exp_setup}(e) and (f). A single dot coupled to one island [Fig.~\ref{fig:exp_setup}(e)] can be used to probe the parity of two Majoranas \cite{Karzig2017}, \emph{i.e.}, a spin expectation value. Moreover, by coupling two dots to Majoranas on neighboring islands as in Fig.~\ref{fig:exp_setup}(f), one can detect the joint parity of the four Majoranas \cite{Karzig2017}, which corresponds to a nearest-neighbor static spin-spin correlation function. 
In Sec.~\ref{sec:ExperiProbes}, we show how dots can also be used to generate Zeeman fields and realize more sophisticated experimental probes.

\subsection{Simplified architectures}\label{sec:SimpArch}

While the architectures described above serve as the elementary building blocks for a large variety of quantum spin models, it may be useful to consider alternative designs with minimal hardware requirements. Such simplified networks can realize only certain models with fewer couplings and may be seen as intermediate steps towards a full quantum simulator described above.

A potential complication of the design in Fig.~\ref{fig:exp_setup} is the noncollinear arrangement of nanowires. This setup precludes applying a magnetic field in the most favorable direction parallel to the nanowire and, hence, perpendicular to the spin orbit direction and parallel to the superconducting film. Moreover, it requires tunnel couplings over longer distances or additional topological nanowires. 

\begin{figure*}[t]
 \includegraphics[width=.8\textwidth]{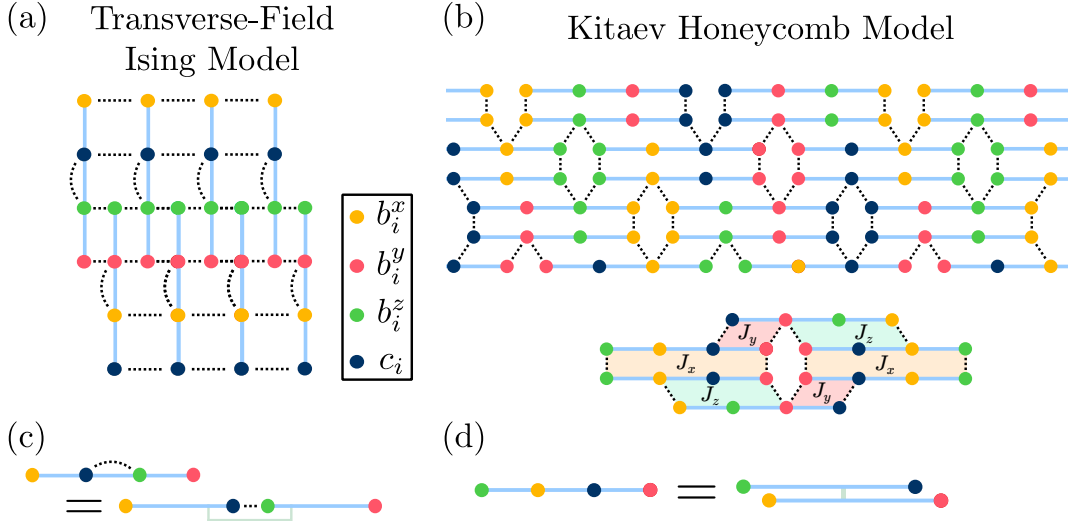}
\caption{Parallel-wire realization of (a) the one dimensional transverse-field Ising model with next-nearest neighbor interactions and (b) the Kitaev honeycomb model. In (b), the direction of the effective exchange interaction between neighboring sites is indicated by the shaded areas. The spin-$1/2$ lattice sites can be constructed from (c) a linear tetron or (d) a two-sided tetron.}
\label{fig:flat_Kitaev}
\end{figure*}

An example of a simplified geometry realizing the one-dimensional transverse-field Ising model is displayed in Fig.~\ref{fig:flat_Kitaev}(a). This is a paradigmatic model for the study of quantum phase transitions and is described by the Hamiltonian
\begin{align}
 H=\sum_i J_i S_i^xS_{i+1}^x+\sum_i h_iS_i^z.\label{TFIM}
\end{align}
The couplings $J_i$ and fields $h_i$ are generated by inter- and intrasite Majorana tunnel couplings. Notably, all wires are parallel, and the tunnel couplings only connect Majoranas over short distances. Each spin site can be made from a single, sufficiently long, proximitized nanowire in the topological phase with an electrostatically depleted central segment, such that domain walls inside the wire harbor two additional Majoranas. The intra-site tunnel couplings required for the Zeeman fields can simply be realized by a finite overlap of Majorana wavefunctions inside the central segment. Alternatively, two topological wires can be coupled by a narrow superconducting bridge to ensure a joint charging energy [see Fig.~\ref{fig:flat_Kitaev}(c)]. In this case tunneling through a quantum point contact between the wires' ends can simulate a Zeeman term.

The model in Eq.~(\ref{TFIM}) is noninteracting and can be used for benchmarking. 
Allowing for additional tunnel couplings between next-nearest neighbors as shown in Fig.~\ref{fig:flat_Kitaev}(a) breaks the integrability of the Ising model and leads to the following Hamiltonian
\begin{align}\label{eqn:2dIsingModel}
 H=\sum_i J_i S_i^xS_{i+1}^x+\sum_i J'_i S_i^xS_{i+2}^x+\sum_i h_iS_i^z.
\end{align}
When the couplings or fields are chosen to be random, this model can be used to study the many-body localization transition. Moreover, it is believed to exhibit a finite temperature spin-glass transition and localization-protected quantum order \cite{Kjall2014}.

A geometry based on parallel wires can even be used to simulate two dimensional models. Figure~\ref{fig:flat_Kitaev}(b) shows a realization of the Kitaev honeycomb model [see Eq.~\eqref{eqn:HKham}], where spin sites are represented by horizontal wires with four Majoranas. This setup again allows for a global magnetic field parallel to the direction of wires and only requires tunnel couplings between Majoranas separated on scales much shorter than the wire length. The lower part of Fig.~\ref{fig:flat_Kitaev}(b) illustrates that the arrangement of tunneling couplings shown indeed realizes Kitaev couplings along the edges of one hexagon. Additional tunnel couplings between Majoranas on the same island can be introduced to add local Zeeman fields.

It is an interesting feature of this geometry that only one of the two inner Majoranas on each site is coupled to a neighboring site and only on side (either up or down). Each spin-$1/2$ site can therefore be realized by two adjacent topological wires as depicted in Fig.~\ref{fig:flat_Kitaev}(d), which minimizes the required size of the superconductor and allows for a sizable charging energy.
We emphasize that the ordering of Majoranas within each islands is important to ensure the mapping to a spin model. In Fig.~\ref{fig:flat_Kitaev} care has been taken to choose networks that map to the Ising and Kitaev models when the fermion parities of all superconducting islands are equal.

\subsection{Implementation}

The physical architecture of our quantum simulator bears many similarities with some of the most promising proposals for a topological quantum computer in Refs.~\citenum{Vijay2015,Landau2016,Litinski2017} and \citenum{Karzig2017,Plugge2017,Vijay2016}. Like our quantum simulator, these quantum computer platforms comprise a network of Majorana islands connected by tunnel couplings whose strengths are greatly exceeded by the charging energy on the island.

While topological quantum computers based on projective measurements only require parity measurement of two or four Majoranas, somewhat more sophisticated measurement protocols are desirable for the quantum simulator (see Sec.~\ref{sec:ExperiProbes}). Nevertheless, external control and characterization of the system can be realized in both cases by employing standard transport measurements and spectroscopy of additional quantum dots.

Despite being based on the same basic building blocks, the proposed quantum-simulator architecture is much less demanding than that of a full Majorana-based quantum computer. Our quantum simulator can therefore also serve as a milestone for experimental efforts geared towards the latter goal. 

\subsubsection{Sign of Heisenberg couplings}\label{sec:ExCouplingSign}

The ability to experimentally control the sign of the tunnel couplings in Eq.~(\ref{eqn:tunHam}) such that the resulting exchange couplings are either ferro- or antiferromagnetic is a key requirement of our proposal. 
Using the low-energy expansion of the local fermion operators discussed in Sec.~\ref{sec:3bond},
\begin{align}
f_{j}^a\simeq u_j^a e^{-i\phi_j/2}e^{i\theta_j^a}b_j^a,
\end{align}
one can write the exchange interaction along the $z$-direction of spins as
\begin{align}
J_{ij}^z=\frac{16\tilde{t}_{ij}^x\tilde{t}_{ij}^y}{E_C} u_i^xu_j^xu_i^yu_j^y{\rm Re}[e^{i(\theta^x_i-\theta^y_i+\theta^y_j-\theta^x_j)}].
\end{align}
by combining Eqs.~(\ref{eqn:HeisenbergCouplingDef}) and (\ref{eqn:t-ttilde}). Because $t^a_{ij}$ and $u^a_j$ are positive numbers the sign of the exchange couplings only depends on the phase differences $\theta_i^x-\theta_i^y$. 

The phases $\theta_i^a$ reflect the Majorana wavefunction at the location of the tunnel coupling. They contain information about the local superconducting phase and the history of creating and braiding Majoranas. When the superconducting gap has a minimum at a finite momentum, the Majorana wavefunction oscillates in space. An experimental consequence of this is the presence of oscillations of the Majorana wavefunction in finite length wires as a function of magnetic field \cite{Albrecht2016}. Due to the small electron density required to enter a topological phase, the wavelength of these oscillations is typically long. In InAs wires the wavelength is $\sim 100 \,$nm. Hence the phase of the Majoranas can be controlled by tuning the distance between the domain wall harboring the Majorana state and the location of the tunnel coupling.

When the band structure has only a minimum at zero momentum, the Majorana wavefunction does not oscillate. In this case, the phases $\theta_j^a$ are entirely determined by the creation and braiding history of the Majoranas and the local superconducting phases. While the overall phase of the superconducting island $\hat\phi_j$ does not assume a particular value in the presence of a charging energy, phase differences between different parts of the same island are well defined as long as Cooper pairs are allowed to move freely within the island. Importantly, the phase of the $p$-wave pairing amplitude depends on the geometric angle between wires \cite{Alicea2011,Halperin2012}, even if the proximity-providing $s$-wave superconductors all have the same phase. 

As an example on how the signs of the exchange couplings can be fixed even when the Majorana wavefunction does not oscillate, we consider the parallel-wire constructions of the Ising and Kitaev models with antiferromagnetic interactions discussed in the previous subsection. Figures~\ref{fig:flat_Kitaev}(c) and (d) show two possible island geometries comprising two parallel topological superconducting wires: the linear tetron and the two-sided tetron \cite{Karzig2017}.
Both wires have the same superconducting phase and the same direction in space. Therefore the two Majoranas localized at the left ends ($b^x_i$ and $b^z_i$ at the bottom of Fig.~\ref{fig:flat_Kitaev}) have identical phases. The phase of the two Majoranas at the right ends ($b^y_i$ and $c_i$) is shifted by $\pi/2$ relative to the left ones. One can readily verify that for these phases, all couplings in the setups displayed in Figs.~\ref{fig:flat_Kitaev}(a) and (b) are antiferromagnetic.

Ensuring the correct signs in models with more couplings and wires at different angles requires a more detailed consideration of the specific experimental setup. In this context, it may be useful to replace certain tunnel couplings between Majoranas by a coupling via a quantum dot. As we show in Sec.~\ref{sec:parity-charge-coupling} below, the sign of such couplings can be tuned by the location of the dot level.

\subsubsection{SU(2) symmetry breaking}\label{sec:SU2sym}

In order to minimize the amount of disorder in the spin model, the tunnel couplings between different islands should be homogeneous across the lattice. This can be approximately realized by a regular design. Moreover, all local couplings can be tuned and probed individually to ensure a minimal amount of fluctuations of the couplings strength between different sites.

An important challenge is to ensure that the SU(2) symmetry of antiferromagnetic Heisenberg models is being realized. This requires equal strength for all three tunnel couplings connecting two islands. Variations between couplings can be eliminated systematically by energy level spectroscopy of two neighbouring sites after temporarily decoupling them from the surrounding islands.
The spectrum of these two antiferromagnetically-coupled sites features three excited states whose energy differences scales with the anisotropy of the interaction. Realizing an SU(2) symmetric Heisenberg term, therefore, simply requires tuning the tunnel couplings to the degeneracy point of the excited states of the two site complex.

This scheme guarantees unwanted SU(2)-symmetry-breaking terms to be bounded by the energy resolution of spectroscopic measurements. Experimental implementations of such measurements are discussed in Sec.~\ref{sec:excitation-spectrum}. While all couplings need to be tuned and probed individually, this step has to be done only once for any particular model realization.  

The degree of SU(2)-symmetry breaking can also affect the effective spin sites constructed from multiple islands depicted in Fig.~\ref{fig:ExtSpins}. Importantly, however, the ground state degeneracy of a spin-$1/2$ site, represented by a chain with an odd number of islands as in Fig~\ref{fig:ExtSpins}(a) remains robust even when the internal exchange interactions are not isotropic. The realization of spins higher than $1/2$ depicted in Fig~\ref{fig:ExtSpins}(b) and (c) requires Heisenberg couplings between several spin-$1/2$ sites. A weak breaking of the SU(2) symmetry of the couplings leads to a small splitting of the spin degeneracy in the case of larger spins.

In principle, SU(2) symmetry can be broken even when all tunnel couplings have identical strengths. This can be traced back to the lack of symmetry between the $b^{x,y,z}$ Majoranas in the tunneling Hamiltonian given by Eq.~(\ref{eqn:tunHam}). As has been discussed in Sec.~\ref{sec:higher-order}, symmetry-breaking terms in the Hamiltonian, however, only occur at higher-order in the small parameter $t^a/E_C$ depending on the length of the shortest loop. In particular, multi-island vertices with higher spin or coordination number remain SU(2) symmetric as long as the tunnel couplings have identical strengths because they generally do not involve any loops.

\section{Experimental Probes}\label{sec:ExperiProbes}

In order to characterize the ground state and excitations of the spin system, external probes are necessary. The effective spin degree of freedom on each site is encoded nonlocally in the parity of two Majorana states. While this makes the spins more robust to decoherence, detecting a spin requires coupling two Majoranas, thereby breaking this protection. Hence, experimental probes of the spin system require some additional architecture.

Measurements of certain local observables can be realized with minor modifications of the setup by adding extra quantum dots tunnel coupled to Majorana states, as shown in Figs.~\ref{fig:exp_setup}(e) and (f). As discussed in the proposals for Majorana-based quantum computers \cite{Karzig2017,Plugge2017}, such dots can be used to measure the two-Majorana parity on a single site, corresponding to a spin expectation value, as well as the parity of four Majoranas on neighboring islands, equivalent to the nearest-neighbor spin correlation function $\braket{\sigma^a_i\sigma^b_{i+1}}$.

Below, we discuss more sophisticated experiments, requiring a relatively limited amount of additional hardware. 
Similar to the parity measurements, we propose to implement experimental control knobs and sensors by coupling the fermion parity of Majoranas to a charge degree freedom. Charges can then be readily read out and manipulated using a variety of all-electric techniques developed for superconducting and semiconducting qubits.

\subsection{Parity-charge coupling}\label{sec:parity-charge-coupling}

In order to measure their parity, two Majoranas need to be coupled such that they form an ordinary complex fermion. This fermion can now have a nonzero energy, as the topological protection of the Majoranas is broken by the coupling. Hence, an unbiased detection of the spin requires the energy of the fermion to be tuned to zero.

While this additional degree of freedom may at first seem like a complication for experiments, we show that it can be exploited to realize more powerful probes of the system. Importantly, a nonzero energy splitting of two coupled Majoranas acts as an effective local Zeeman field in the spin model. As we shall see below, the same architecture used for measuring spins can also be used as a knob to generate excitations, thereby giving access to dynamical properties of the system.

In the simplest case, a coupling between two Majoranas can be realized by introducing a direct tunnel coupling between them. The sign and magnitude of the coupling term, however, depends on the details of the Majorana wavefunction and is difficult to tune dynamically.
Therefore, we instead propose that a versatile two-Majorana coupling may be induced by allowing tunneling to an additional quantum dot or by introducing a different charging energy for a part of the Majorana island.

\subsubsection{Additional quantum dots}\label{sec:add-quantum-dots}

Experimental control of the Majorana parity can be obtained by exploiting the fact that the energy levels of a quantum dot tunnel coupled to two Majoranas on a single superconducting island depend on the joint Majorana parity.
This dependence originates from virtual processes where an electron from the dot tunnels into one Majorana and onto the superconducting island and back to the dot via the other Majorana. This effect has been proposed previously as a tool for readout measurements of a Majorana qubit \cite{Karzig2017,Plugge2017}. Here, we propose an modified setup that also yields access to dynamic properties of the system.

\begin{figure*}[t]
 \includegraphics[width=.98\textwidth]{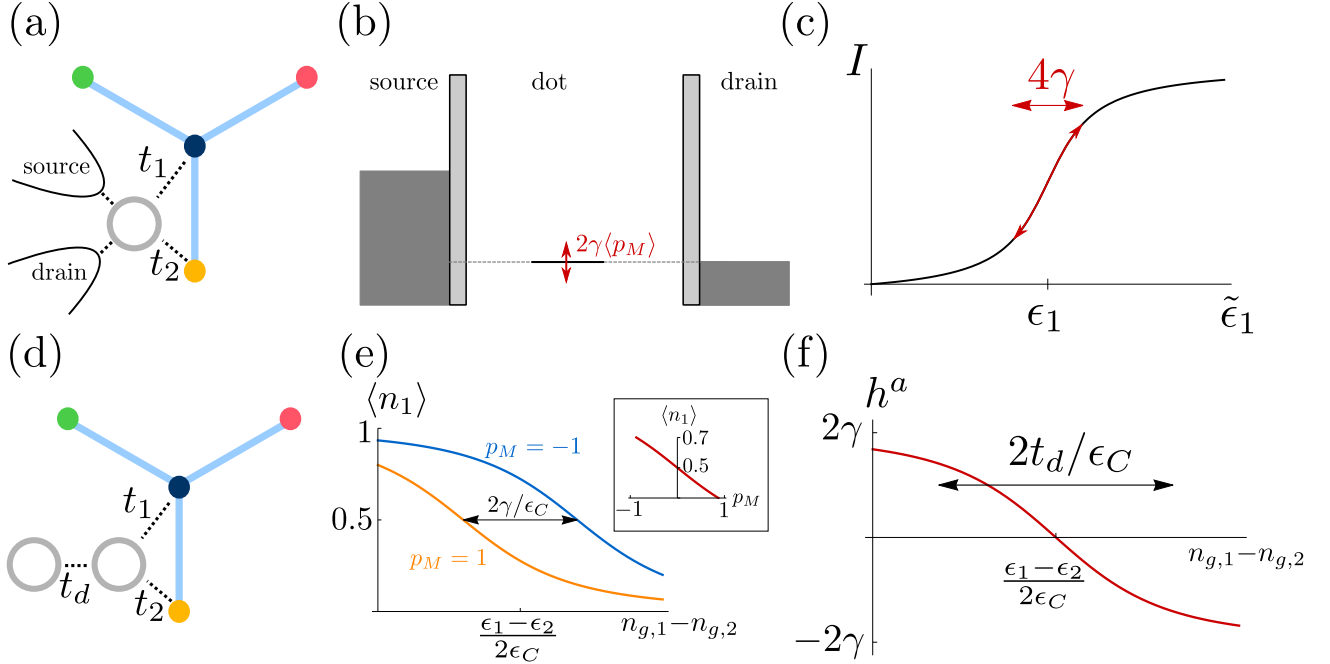}
\caption{(a) Setup for parity measurements using a single quantum dot coupled to two Majoranas. External leads are used for transport measurements through the dot. (b) Energy diagram of the dot. When the dot level is close to the chemical potential in one of the leads, the current shown in (c) changes rapidly with the level position $\tilde\epsilon_1$.
(d) Setup for parity measurements using a double dot with one dot tunnel coupled to two Majoranas on the same superconducting island. (e) Charge on the first dot as a function of gate voltage for two different Majorana parity states $p_M=\pm 1$ at zero temperature and $\gamma=t_d/2$. The inset shows the charge on the dot as a function of the Majorana parity at the degeneracy point $n_{g,1}-n_{g,2}=(\epsilon_1-\epsilon_2)/2\epsilon_C$.
(f) Effective Zeeman field strength $h^a_i$ equal to the the energy difference between the ground state energies of the coupled Majorana-dot system for both even and odd Majorana parity and $t_d=8\gamma$.}
\label{fig:dot}
\end{figure*}

We consider a dot whose charging energy $\epsilon_C$ and level spacing greatly exceed temperature so that it is sufficient to focus on a single electronic level with orbital energy $\epsilon_1$. We can then express the dot Hamiltonian in terms of the dot level creation operator $d_1$ as
\begin{align}
H_{1}=\epsilon_1d_1^\dag d_1+\epsilon_C(d_1^\dag d_1-n_{g,1})^2,
\end{align}
where $n_{g,1}$ is the charge induced by a gate voltage. The dot is operated in the regime near the degeneracy point where the filled and empty state have equal energy, $n_{g,1}\simeq (\epsilon_1+\epsilon_C)/2\epsilon_C$.

We allow for tunneling between the dot and two Majoranas $b_{1,2}$ on the same superconducting island as shown in Fig.~\ref{fig:dot}(a). The tunneling can be described by the Hamiltonian
\begin{align}
H_{t} =(t_1 b_1+t_2 b_2)d_1^\dag e^{-i\phi/2} +{\rm h.c.}
\end{align}
Because the superconductor has a large charging energy, $E_C$, only virtual tunneling events are possible. To leading order in $1/E_C$ we obtain $H=H_{1}+H_{t,{\rm eff}}$
\begin{align}
H_{t,{\rm eff}} =\gamma ib_1b_2 (2d_1^\dag d_1-1)
\end{align}
where we have introduced the coupling strength $\gamma=2{\rm Im}(t_1^*t_2)/E_C$. The effective dot Hamiltonian assumes the form $H_1$ with the replacement $\epsilon_1\to \tilde\epsilon_1=\epsilon_1+2\gamma ib_1b_2$. The dot levels therefore acquire an energy shift that is proportional to the Majorana parity $p_M=ib_1b_2$, allowing one to to determine the spin expectation values $\Braket{p_M}$ through a simple measurement of the energy shift.

The energy shift can be observed in transport measurement by passing a current through the dot in the Coulomb blockade regime. When the dot is tuned to the edge of a Coulomb diamond at finite bias, i.e,  the sum of the dot level and the charging energy becomes degenerate with the Fermi level in one of the leads [see energy diagram in Fig.~\ref{fig:dot}(b)], the current depends sensitively on the dot level position. If the coupling strength $\gamma$ between the dot and the Majoranas is smaller than the tunneling rate, the current varies approximately linearly with the Majorana parity as shown in Fig.~\ref{fig:dot}(c).

The Majorana parity can alternatively be determined by a charge measurement by coupling a second quantum dot or an external lead to the first dot. Tuning the dot energy level to a degeneracy point of the coupled system enables the dot to fluctuate, and the parity dependent shift of the dot level then translates into a parity dependent charge. To be specific, we consider a double dot setup containing a single electron, as shown in Fig.~\ref{fig:dot}(d) (similar arguments apply to an external lead). The Hamiltonian acquires an additional term
\begin{align}
H_{12}=\epsilon_2d_2^\dag d_2+\epsilon_C(d_2^\dag d_2-n_{g,2})^2+t_d(d_1^\dag d_2 +d_2^\dag d_1),
\end{align}
where the operator $d_2$ describes the fermion on the second dot. When the dot levels are detuned from each other, the charge is localized predominantly in one of the two dots. By varying the gate potentials $n_{g,1/2}$, the charge distribution can be smoothly moved to the other dot. In the absence of the coupling to the Majoranas, $t_1=t_2=0$, the charge is equally distributed between the two dots at the degeneracy point $n_{g,2}-n_{g,1}=(\epsilon_2-\epsilon_1)/2\epsilon_C$.

We now consider the effect of the coupling $H_t$ between the Majoranas and the first dot. For simplicity, we assume $E_C$ to be much larger than the energy required to change the charge on the double dot system, although an extension to the more general case is straightforward \footnote{The double dot states $(1,1)$ and $(0,0)$ differ in energy from the states $(0,1)$ and $(0,1)$ because of the cross capacitance between the dots.}. In this case, the effective Hamiltonian simply reads $H=H_1+H_{12}+H_{t,{\rm eff}}$, and after projection to the single-charge sector, we obtain 
\begin{align}
H=\epsilon_C( n_{g,2}^2+n_{g,1}^2)+\begin{pmatrix}\delta_1+\gamma p_M&t_d\\
t_d&\delta_2-\gamma p_M
\end{pmatrix}
\end{align}
in the basis $\{\ket{10}, \ket{01}\}$, which comprises the Fock states of the two dots with a total filling of one. Here, $p_M=ib_1b_2$ is the effective spin operator defined by the two Majoranas, and we have used the short-hand notation $\delta_i=\epsilon_i-2\epsilon_C n_{g,i}$.

As above, the coupling to the Majorana simply shifts the energy level of the first dot in a parity dependent way. Close to the degeneracy point of the two dots, this leads to charge accumulation or depletion on the first dot depending on the sign of the shift, and it follows that the charge on the first dot depends on the parity [see Fig.~\ref{fig:dot}(e)]. To determine the parity by a charge measurement, the dot should ideally be in the regime $t_d\gtrsim \gamma$ since this is where the relation between dot charge and Majorana parity at zero temperature is approximately linear, as shown in the inset of Fig.~\ref{fig:dot}(e). Moreover, the coupling $\gamma$ should be small compared to the energy scales of the spin Hamiltonian, such that the latter is only minimally perturbed.

At small but finite temperatures, the first excited state of the double dot is thermally occupied near the degeneracy point, where the excitation energy is minimal. Hence, the effect of nonzero temperature is similar to that of tunneling between the dots: the charge varies continuously from 0 to 1 when the gate voltages are tuned across the degeneracy point.

The setup described above can also be operated in reverse: because one parity state is favored energetically, one can generate an effective Zeeman field by tuning the gate voltage on the dot. In this case, the most favorable regime is $\gamma\ll t_d$, where the energy difference between the two parity states (i.e., the Zeeman field strength) is linear in the gate voltage around the degeneracy point [see Fig.~\ref{fig:dot}(f)].

It is also possible to work in the opposite regime $\gamma\gg t_d$, where the relationship between the effective Zeeman field and the gate voltage is again linear. One should keep in mind, however, that the coupling strength $t_d$ also determines the gap in the double quantum dot. To minimize the effect of the probe on the spin system, $t_d$ should be adjusted so that is is larger than the typical energy scale of the simulated spin Hamiltonian. This is particularly important when the quantum dot is used to probe the excitation spectrum or the dynamic structure factor of the spin system as discussed below. Such measurements require the probe frequency to remain below the gap of the double dot.

\subsubsection{Split Majorana island}\label{sec:split-island}

\begin{figure}[t]
 \includegraphics[width=.47\textwidth]{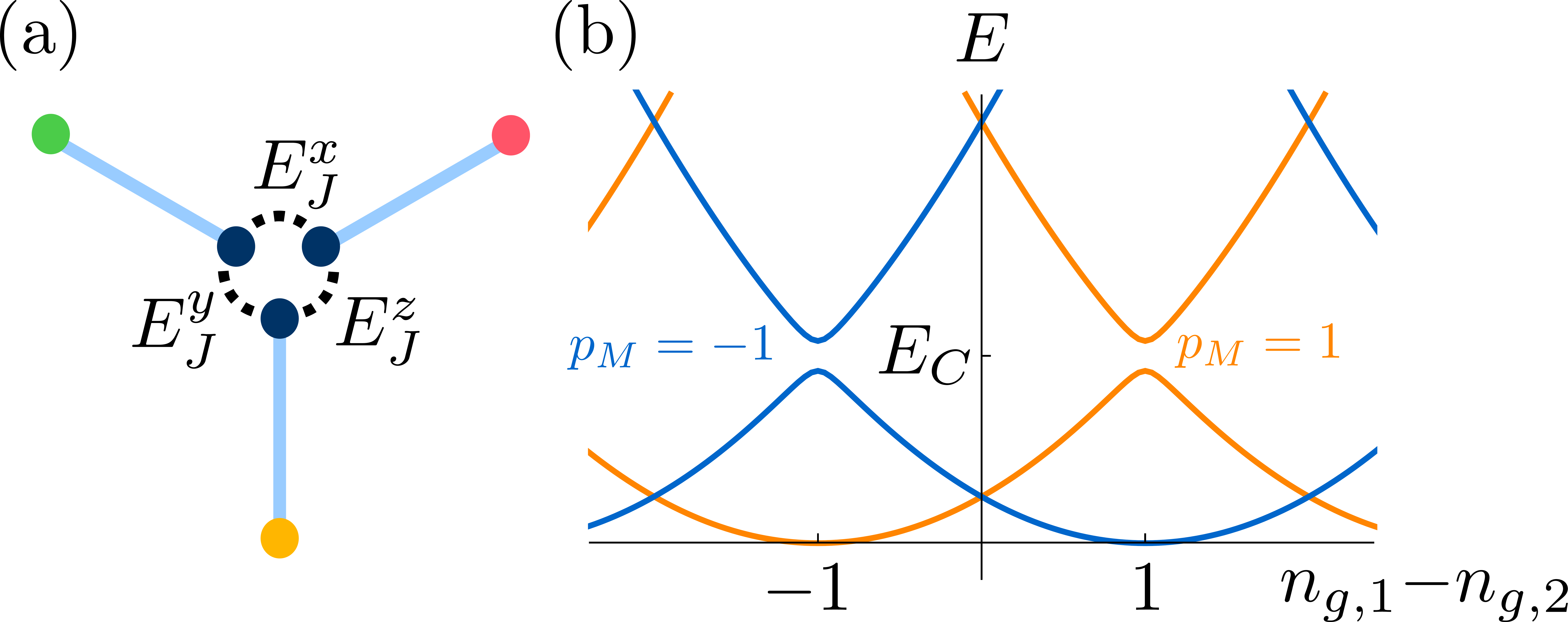}
\caption{(a) Modified setup of the Majorana island with three wires coupled by Josephson junctions. For a choice of couplings $E_J^x\gg E_C$ and $E_J^{y,z}\sim E_C$, the island decomposes into two subislands, whose charge parity is associated with the expectation value of the spin operator $\sigma_x$.
(b) Spectrum of the Majorana island as a function of the difference of gate voltages $n_{g,1/2}$ on the two subislands (see Ref.~\onlinecite{Barkeshli15}). The Josephson couplings are $E_J^{y,z}=E_C/50$ and $E_J^x=\infty$.}
\label{fig:split_island}
\end{figure}

As an alternative realization of a parity-charge coupling, we propose a slightly modified design of the Majorana island, in which three topological wires with separate charging energies are mutually coupled at their ends by Josephson junctions (see Fig.~\ref{fig:split_island}). The junctions can be implemented by simply adding gate controlled barriers to the previously discussed setup. When all Josephson couplings $E_J^a$ greatly exceed the charging energies, the charge gap of the individual islands are exponentially suppressed, and this setup simply realizes the Majorana island with a global charge constraint and a twofold degenerate ground state.

When one coupling $E_J^x$ remains very strong while $E_J^{y,z}$ are reduced to a scale of order $E_C$, 
the ground state degeneracy is lifted. The Majorana island decomposes into two subislands, where the wire hosting the $b_i^x$ Majorana is only partially coupled to the other two wires. As long as the imbalance of gate-induced charges on the two subislands is not too large, the two low-energy states remain approximately degenerate and well separated from all higher energy states by a gap $\sim E_C$ [see Fig.~\ref{fig:split_island}(b)]. The two low-energy states correspond to the subisland's fermion parities states and hence the two eigenstates of the spin operator $\sigma^x_i$. 

This setup therefore effectively realizes a spin-$1/2$ in the presence of a Zeeman field. The Zeeman field depends on the strength of the Josephson couplings relative to the charging energy and can be tuned to zero by varying a local gate potential, restoring the groundstate degeneracy. An exact expression for the Zeeman field can be evaluated in terms of Mathieu functions \cite{Koch2007} and has been given in Ref.~\citenum{Barkeshli15}.

The Majorana parity can be measured with a nearby charge detector that is capacitively coupled to either subisland. Such a measurement is possible when Cooper pair tunneling between the subislands is weak, $E_J^{y,z}\ll E_C$, and the two low-energy eigenstates near $n_{g,1}-n_{g,2}=0$ have well-defined charges which differ by one. In this limit, the charge expectation value can be associated with the expectation value of the  Majorana parity operator.
The couplings $E_J^y$ and $E_J^z$ can in principle be chosen arbitrarily small. In models where the $c$-Majorana must couple with all of its neighbours, however, it is important to ensure that it remain delocalized over all three wires in Fig.~\ref{sec:split-island}(a).

The setup in Fig.~\ref{fig:split_island}(a) is related to the proposed realization of a spin-$1/2$ site in Ref.~\onlinecite{Barkeshli15}. Using this setup as an elementary building block for a lattice, however, has the drawback that the Zeeman term must be fine-tuned to zero. This makes each spin susceptible to charge noise in the gate, which acts as a fluctuating effective magnetic field in the spin model. It is instead advantageous to choose strong Josephson couplings $E_J^a\gg E_C$ on all sites except for one site used for measurements. In this way, the spin system remains largely insensitive to noise.

\subsection{Experimental observables}

While simple charge measurements can determine the spin expectation values at each site, the probes introduced above can also provide access to key properties of the spin system such as the spectrum of spin excitations and the spin-spin correlation functions. Finding the excitation spectrum requires a local measurement of a quantum dot spectrum, which has been realized previously in a variety of experiments. The correlation function provides more comprehensive information, but is also more challenging as it requires nonlocal measurements. We propose several experiments for both quantities.

\subsubsection{Excitation spectrum}\label{sec:excitation-spectrum}

The density of states of a single Majorana island at fixed charge corresponds to the local density of states of the simulated spin model. In the majority of cases, the most relevant information is the spectrum of extended spin excitations, and this can be obtained by a local measurement on a single site. When the spin system has localized modes, sampling the measurement over various spin sites may be necessary to ensure  sizable transition matrix elements to all states.

The spectrum of different charge states of a quantum dot can be detected by charge reflectometry \cite{Majer2007}, which requires a capacitive coupling between the subisland or quantum dot and a transmission line cavity (see Ref.~\onlinecite{Frey2012} for an example of a double dot coupled to a cavity). The absorption of the cavity mode is strongly enhanced when its frequency matches the transition energies between dot levels of different charge.

Here, we are interested in excitations within a fixed-charge sector on each individual island. Charge reflectometry of the entire island, however, would probe states outside the Hilbert space of the spin system. Instead, the transmission line resonator needs to be coupled to a dot (or subisland) whose charge is coupled to a Majorana parity as described in Sec.~\ref{sec:parity-charge-coupling}. The spectrum can be measured by detecting absorption of microwave photons inside the cavity. If the system is tuned to the degeneracy point where the Zeeman field vanishes, the coupling of the site to the cavity mode can be described by the Jaynes-Cummings Hamiltonian
\begin{align}
    H=H_{\rm spin}+\beta \sigma_i^a(a+a^\dag)+\omega_r\bigr(a^\dag a+\frac{1}{2}\bigr),
\end{align}
where the coupling strength $\beta$ depends on the properties of the cavity and the capacitance between the cavity and the subisland \cite{Blais2004} and $\sigma_i^a$ is the effective spin defined by the Majorana parity. Absorption of cavity photons results in spin flips and hence induces transitions between the many-body states of the spin systems. The excitation spectrum of the spin system can therefore be observed by detecting resonances in the cavity transmission as long as the operator $\sigma^a_i$ is not a conserved quantity of the Hamiltonian.

\begin{figure}[t]
 \includegraphics[width=.45\textwidth]{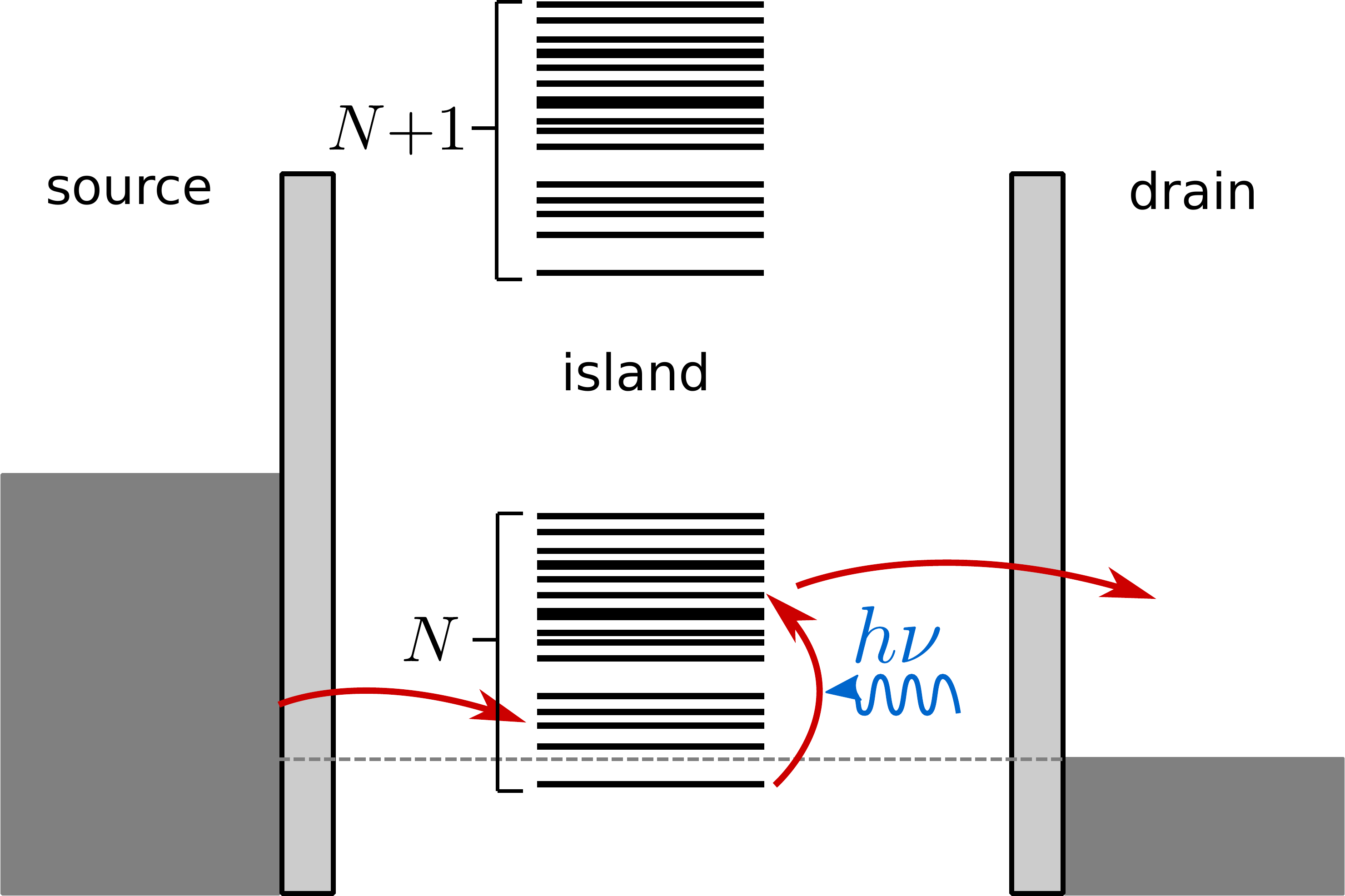}
\caption{Transport experiment to measure the spectrum of the spin system. An ac gate voltage applied to a subisland results in transitions to excited states within the same charge sector, which can then decay into the drain. The island with one missing electron is subsequently more likely to be refilled from the source lead.}
\label{fig:spin_blockade}
\end{figure}

Alternatively, the spectrum can be probed in a transport experiment in the Coulomb blockade regime.  Two Majoranas on the same island are contacted by leads at different voltages. The gate voltage of the island is chosen such that the groundstate of the spin system is stable, while allowing higher energy excitations to decay into the drain  by transferring an electron from the dot to the lead. This can be achieved by choosing a gate voltage slightly below the energy threshold for removing an electron out of the equilibrium state, but high enough that electrons can escape when the spin system is in an excited states [see Fig.~\ref{fig:spin_blockade}]. The voltage in the source lead is chosen sufficiently high such that no spin excitation can decay into the source lead, and sufficiently low that higher charge states of the superconducting island remain unoccupied.

Since all excited states can decay, the system will eventually relax to the ground state, where transport is blocked. 
By applying an alternating gate voltage to the subisland (or the quantum dot coupled to the Majoranas) one can generate an ac Zeeman field, which induces transitions to excited states of the spin system. Electrons from the island can then tunnel into the drain, leaving the Majorana island  with one missing electron. This is then filled with an electron from either lead by another tunneling event. If the final state is again an excited state, transport continues until the system reaches the ground state.

By applying periodic ac pulses to the gate, a steady state current can be measured whenever the ac frequency matches a transition between the ground state and an excited state. A similar experiment has been performed in the context of spin blockade in semiconductor quantum dots \cite{Nadj-Perge2010}. 
To avoid transitions between excited states due to the external radiation, the period should exceed the relaxation time to the ground state.

If the excitation spectrum consists of many states, however, the probability of ending up in the ground state after the dot is refilled can be very low, resulting in a long relaxation time. The relaxation time can be reduced by choosing asymmetric tunnel couplings such that refilling the dot from the drain side becomes more efficient. Alternatively, the bias voltage in the source lead can be reduced so that only low-lying excited states are energetically accessible for source electrons.

\subsubsection{Measuring correlation functions}\label{sec:CorrFunMeas}

More comprehensive information about the system can obtained from frequency-dependent spin-spin correlation functions. A key observable for the characterization of the spin model is the retarded spin-spin correlation function
\begin{align}
 \chi_{ij}^{ab}(t-t')=-i\Braket{[\s^a_i(t),\s^b_j(t')]}\th(t-t'),
\end{align}
which governs the response of a spin to a time dependent magnetic field according to
\begin{align}
 \Braket{\delta \s_i^a(t)}&=\int_{-\infty}^\infty dt'\sum_{j,b} \chi^{ab}_{ij}(t-t';j\ell)h_j^b(t')
\end{align}
with $\delta \sigma^a_i(t)=\sigma^a_i(t)-\braket{\sigma^a_i}$, or in Fourier domain
\begin{align}
  \Braket{\delta \s^a(k,\omega)}&=\sum_{b} \chi^{ab}(k,\omega)h^b(k,\omega).\label{linear_response}
\end{align}
In the context of spin models, it is more common to consider the dynamic structure factor
\begin{align}\label{eqn:DynStucFactor}
S^{ab}(k,\omega)=\sum_{j}\int d t \,e^{i\omega t-i kj} \braket{\delta \sigma^a_j(t)\delta\sigma^b_0}.
\end{align}
The correlators are related by the fluctuation-dissipation theorem \cite{landau1980}
\begin{align}
 S(k,\omega)=\frac{2}{1-e^{-\beta \omega}} {\rm Im}\chi(k,\omega)\label{fluctuation_dissipation}
\end{align}
where $\beta$ is the inverse temperature.

The frequency dependent response function can be determined by measuring the nonlocal spin response to a locally applied external field. By applying an oscillatory Zeeman field $h^b_j(\omega)$ on site $j$ and measuring the expectation value of a spin $\braket{\sigma^a_i(\omega)}$ on a different site $i$, one can directly obtain the correlation function $\chi^{ab}_{ij}(\omega)$ via Eq.~(\ref{linear_response}). 

An oscillatory Zeeman field can be realized by applying an ac gate voltage to a quantum dot or subisland whose charge is coupled to the Majorana parity as discussed in Sec.~\ref{sec:parity-charge-coupling}. Measuring the expectation value $\braket{\sigma^a_i(\omega)}$ requires a time-dependent parity measurement of two Majoranas. This can be achieved, for instance, by weakly coupling a quantum dot to the pair of Majoranas. As explained in Sec.~\ref{sec:add-quantum-dots}, this setup can be tuned into a regime where the conductance through the dot varies with the Majorana parity and thus with $\braket{\sigma^a_i}$ \cite{Karzig2017}.  It follows that finite-frequency oscillations of the expectation value $\braket{\sigma^a_i(t)}$ result in an alternating current at the same frequency. 

Measuring $\braket{\sigma^a_i(\omega)}$ as a function of drive frequency and position yields access to the full energy- and momentum-dependent response function. The complex phase of $\chi^{ab}_{ij}$ can be obtained from the phase mismatch between the applied ac voltage and the detected dc current, which is accessible via lock-in techniques. Moreover, one can determine the dynamic structure factor from the response function using Eq.~(\ref{fluctuation_dissipation}). 

The response function can alternatively be accessed by coupling subislands (or nearby quantum dots) on two spin sites to transmission line resonators as in Sec.~\ref{sec:excitation-spectrum}. By driving one spin at the probe frequency and measuring the transmission and phase shift in the resonator coupled to the other spin site, one can directly obtain the response function.

An interesting question of practical relevance is whether the correlation function can also be measured with a single resonator coupled to two different spins. Nonlocal spin-spin correlations should result in a beating pattern in the resonator signal, however, extracting a correlation function from the reflection properties of a single  resonator requires a more detailed analysis of the coupled resonator-spin system. Alternatively, a measurement of the correlation function with a single resonator could be realized by a pump-probe experiment, where the couplings between the resonator and the spins are switched on and off at different times. The switching of the resonator couplings can be realized by rapidly detuning the quantum dots that mediate the charge-parity couplings.

\subsection{Experimental parameters}\label{sec:ExpParam}

To judge the feasibility of these proposals, we now estimate experimental parameters specific to networks of semiconductor nanowires with proximity induced superconductivity. The typical charging energy of an InAs wire with a thin Al shell \cite{Albrecht2016}, $E_C\simeq 160\, \mu$eV, is of similar magnitude as the induced gap, $\Delta\simeq 180 \, \mu$eV. The tunneling strength between two Majoranas can be arbitrarily tuned by a gate voltage. By choosing $t\simeq 0.3 E_C$, we obtain an exchange interaction $J\sim 4t^2/E_C= 0.36 E_C=60\, \mu$eV. This exceeds the typical electron temperature of $T=50\,$mK (corresponding to $4\,\mu$eV) \cite{Albrecht2016} by an order of magnitude. Alternatively, relaxation of qubits via a coupling to the transmission line resonator can yield a temperature as low as $T=35\,$mK \cite{Jin2015}.

With this choice of parameters, the leading corrections to most spin Hamiltonians are of order $t^2/E_C^2\simeq 10\%$, as already mentioned in Sec.~\ref{sec:higher-order}. 
If a higher accuracy is needed, the tunnel couplings can be adjusted accordingly.

The quasiparticle poisoning time of the Majorana islands sets a limit on the duration of experiments that require a coherent time evolution.
This is particularly relevant for understanding the dynamical properties of isolated systems. 
The poisoning time of a Majorana qubit on an island that is not connected to external leads can be of the order of $\tau_\mathrm{qpp}\sim10\,$ms \cite{Higginbotham2015,Albrecht2017}, and in future devices potentially even considerably higher \cite{Woerkom2015}.
However, the current estimates are more than sufficient when compared against the fastest timescale of the effective spin system, $\tau_J=h/J\sim 60\,$ps, implying that $\tau_\mathrm{qpp}\sim 10^7 \tau_J$.

The observation of dynamic quantities requires measurements in the frequency regime $0.1-10\,$GHz  (set by the exchange interaction), which is well within the capabilities of standard experiments. We expect the frequency resolution of the photon-assisted transport experiment proposed in Sec.~\ref{sec:excitation-spectrum} to be similar to the spin blockade experiment in Ref.~\onlinecite{Nadj-Perge2010}, where the resolution is $\Delta f\sim 0.2\,$GHz. Considerably enhanced resolution is offered by spectroscopy with a transmission line resonator with $\Delta f\sim 2\,$MHz \cite{Majer2007,Frey2012}.

Measurements in the time domain are also feasible. Rabi oscillations have been observed in semiconductor quantum dots with frequencies in the relevant range of $\sim 10\,$MHz \cite{Petta05,Nadj-Perge2010} to $\sim 10\,$GHz \cite{Kim2014,Kim2015}. 
These measurements are particularly relevant to our quantum simulator as they perform continuous measurements of the occupation probability rather than individual projective measurements. 
For the simulation of quenches (see Sec.~\ref{sec:DynProp} below), reasonably short rise times for voltage pulses are necessary. For the control of a superconducting qubit in a parameter regime relevant to our proposal, a voltage rise time of $\sim 30-40 \,$ps has been reportd \cite{Nakamura1999}.
In comparison, the fastest timescale of the effective spin system is $\tau_J= h/J\sim 60$ ps.

\section{Applications}\label{sec:Applications}

Having established that the properties of the Majorana island networks we propose are governed by effective spin Hamiltonians, we now discuss some potential applications of this framework.
Our primary goal is to underscore the versatility and usefulness of our Majorana simulator.

Our simulator is perhaps most obviously applicable to the field of frustrated magnetism, especially in two dimensions.
Nevertheless, it's important to acknowledge that spin models in one dimension require significantly less hardware, and their simulation can be a useful milestone for experiments aiming to realize more complex systems.
Similarly, the architecture proposed in Sec.~\ref{sec:SimpArch} considerably simplifies the construction of certain models. While we equally mention both difficult, far-future projects and simpler, more experimentally realizable applications, we emphasize the latter when appropriate.

This section is divided into several broad (and occasionally overlapping) categories.
We start by discussing the identification of groundstates and the characterization of their properties before turning to applications to out-of-equilibrium systems.

\subsection{Groundstate properties}\label{sec:ApplicGndState}

The quantum simulator can be used to study groundstate properties of spin Hamiltonians in one and two dimensions. Some example models have been provided in Sec.~\ref{sec:HamExamples} and we now expand on these remarks and further motivate their study.

Entering the equilibrium of a realistic model should be straightforward.
Immediately after the tunneling amplitudes between Majoranas have been tuned to their desired values, the network will realize a generic superposition of energy eigenstates of the final Hamiltonian.
The details of this initial state are not expected to be important since the groundstate is subsequently attained through relaxation.
This may appear difficult to achieve given the protection the Majorana islands provide against decoherence; however, we claim that this is actually an advantage, as it leaves the experimentalist largely in control of the relaxation process.
For instance, dissipation and temperature may be adjusted by allowing the system to interact with external leads.
At leading order in perturbation theory, this is equivalent to coupling the effective spins of the islands to external spin operators composed of conduction electrons in the lead \cite{Beri2012}.
For island $i$, this interaction is
\begin{align}\label{eqn:KondoCoupling}
    H_{\mathrm{bath},i}&= \sum_{a=x,y,z} \lam^{a}_i \s_i^a L_i^a.
\end{align}
Here, $L_i^a=\sum_{b,c}\ep^{abc}\psi_{b,i}^\dag\psi_{c,i}$ is a spin-1 vector where $\psi_{a,i}$ is the conduction electron which couples to the $b^a_i$ Majorana.
The coupling strength is $\lam_i^a\sim\sum_{b,c}\abs{\ep^{abc}}\eta_i^{b}\eta_i^{c}/E_C$ where $E_C$ is the charging energy and $\eta_i^a$, $a=x,y,z$ are the (small) tunneling amplitudes of the Majoranas and the conduction electrons of the leads.
Both the temperature and density of states of the leads can be manipulated to allow for the desired relaxation rate.
Provided the temperature scale and Majorana-lead coupling strength are sufficiently smaller than the charging energy and the inter-spin exchange respectively, the series of leads should act as an external bath for the Majorana network.

Alternatively, instead of coupling the Majoranas directly to leads, we could use the hardware for the measurement protocols outlined in the previous section.
There, the islands couple exclusively to quantum dots, which in turn couple to external leads or transmission line resonators.
In this scenario, the latter would act as the bath.

Regardless of the relaxation process, the lowest temperature scale that can be reached is set by the physical temperature of the setup, $\sim 35-50$ mK \cite{Albrecht2016,Jin2015}.
Relative to the typical value of $J$ [see Sec.~\ref{sec:ExpParam}], this implies $T/J\sim0.05-0.07$.
It is reasonable to assume that this is small enough that the equilibrium state reproduced by the simulator accurately reflects the groundstate of the system.

We now compare the observables accessible in our simulator against those obtained in experiments. 
We next present some specific problems our simulator can address, focusing on frustrated magnetic systems with and without disorder, quantum phase transitions, and the entanglement entropy.

\subsubsection{Comparison against experimental observables}

Since the end-goal of a quantum simulator is to improve our understanding of physical materials, it is important to compare the data our simulator can provide against the information obtained using current experimental techniques. 

In Sec.~\ref{sec:ExperiProbes}, we demonstrated that the onsite magnetization $\Braket{\v{S}_i}$ can be directly measured.
This makes the detection of magnetic order relatively straightforward.
By connecting the system to leads, $\Braket{\S_i}$ may be tracked as a function of temperature. 

We also showed how the spectral function and the dynamic structure factor $S^{ab}(x_i-x_j,t)$ defined in Eq.~\eqref{eqn:DynStucFactor} can be obtained through linear response. 
This data probably most resembles the information obtained through inelastic neutron scattering (INS).
An INS measurement proceeds by sending a beam of neutrons towards the sample.
There, the neutron spin interacts with the spins of the material.
Based on the momentum of the neutrons exiting the sample, one can deduce the structure factor $S^{ab}(k,\omega)$ of the sample, where $k$ is the momentum and $\omega$ the energy.
It follows that a direct comparison between our results and those of INS requires a momentum sum. 
We note that neutrons also couple to the phonon modes of the crystal, and that this can disrupt the signal and make determining the spin contribution difficult. 
The large size of the Majorana islands precludes such a problem from occurring within our simulation platform.
Because the neutron beam cannot be perfectly collimated, another issue is that it may be difficult to resolve $S^{ab}(k,\omega)$ at low energies and momenta.
Conversely, the resolution of Majorana network is limited only by its size.

Finally, integrating the structure factor over all of space returns the magnetic susceptibility. 
This can be tracked as function of temperature to extract the Curie temperature.

\subsubsection{Magnetic systems}\label{sec:FrustMag}

The Heisenberg model is the canonical model describing magnetic systems.
While the groundstate properties of many Heisenberg model variants are well-established, many others remain poorly understood.
In particular, our understanding of  magnetic systems with significant frustration is far from complete. 
These systems may not develop magnetic order, even at low temperatures, and one especially fascinating non-magnetic alternative is an exotic quantum phase of matter, typically called a ``spin liquid"  \cite{savary17,SLreviewZhou17,mendels11,NormanKagome16}.
The frustration can originate either from the geometry of the lattice, as is the case with the kagome antiferromagnet, or from the couplings themselves, as we saw with the Kitaev-type models.
Identifying and studying these phases is often challenging for a number of reasons. 
First, obtaining experimental measurements of candidate materials and interpreting the results can be quite difficult.
For instance, while the absence of magnetic order is one hallmark of a spin liquid, it is not a sufficient diagnostic. 
Their true defining feature is the presence of nonlocal excitations, and these are very difficult to identify experimentally.
With the control afforded by our setup, anyonic excitation could be seeded as defects, and possibly even transported and braided \cite{Barkeshli15,Nayak08}.
Alternatively, when anyonic excitations are present, spectroscopic measurements should have a universal form \cite{Morampudi17}.
Without phonons or impurities to undermine the signal, this may be easier to observe in our setup than in a physical system.

Another issue is that even simplified models can be difficult to investigate numerically, especially those for which exotic phases are expected to appear.
In particular, the antiferromagnetic Heisenberg model on non-bipartite lattices has a ``sign problem," which precludes the use of quantum Monte Carlo (QMC) \cite{Lauchli11,Troyer05}.
A myriad of other computational techniques have been developed, but system size quickly becomes a limiting factor (in dimensions greater than one), leaving a fair amount of room for ambiguity.

The spin-1/2  antiferromagnetic Heisenberg model on the kagome lattice is one model of notable interest \cite{mendels11,NormanKagome16}.
Despite having been a topic of study for several decades, its groundstate remains disputed \cite{kagomeDMRGYan,kagomeDMRGDependbrock,Mei17tensor,Jiang12,iqbal11,iqbal13,iqbal15,kagomeZaletel17,kagomeLauchli16}. 
At this time, there are two primary proposals: the $\mathds{Z}_2$ spin liquid \cite{sachdev92} and the Dirac spin liquid \cite{hastings00,ran07,hermele08}.
The essential difference between the two theories is the presence or absence of a spin gap.
Ascertaining the groundstate is meaningful not only from a theoretical point of view, but also experimentally, as the kagome Heisenberg model is believed to describe the compound Herbertsmithite (ZnCu$_3$(OH)$_6$Cl$_2$) \cite{Herbersmithite}.
In Fig.~\ref{fig:4bond}(c), we show a potential setup which could be used to simulate this Hamiltonian.
If present, the energy gap is estimated to be in the range $0.05-0.1J$ \cite{mendels11,kagomeDMRGDependbrock,Nishimoto13}, and with $T/J\sim0.05-0.07$, signatures of a gap should be observable via the techniques discussed in Sec.~\ref{sec:ExperiProbes}.
Further, we emphasize that our estimate of $T/J$ is based on current experimental parameters, and these may improve under future developments.

The XXZ model on the kagome lattice has also attracted recent interest. 
It was realized that the point with $\cJ^z/\cJ^{xy}=-1/2$ is exactly solvable and possesses an exponentially large manifold of exact groundstates \cite{changlani18}.
Further, analoguous exactly-solvable points actually exist for other XXZ models on lattices composed of triangles, such as the triangular and Shastry-Sutherland lattices.

Another class of frustrated Hamiltonians worth exploring are the Heisenberg-Kitaev Hamiltonians,  given in Eq.~\eqref{eqn:HKham} \cite{KHmodel_Chaloupka10}.
The honeycomb lattice version is believed to be relevant for the iridate oxide compounds, Na$_2$IrO$_3$ and Li$_2$IrO$_3$ \cite{PhysRevB.82.064412,PhysRevB.83.220403,PhysRevLett.108.127203,PhysRevLett.108.127204,PhysRevB.85.180403,PhysRevLett.109.266406,PhysRevLett.110.076402,PhysRevLett.111.037205,PhysRevLett.110.097204}. 
By changing $\cJ_H/\cJ_K$, three phases should be accessible: a N\'{e}el antiferromagnet, the stripy antiferromagnet, and the spin liquid \cite{KHmodel_Chaloupka10,PhysRevB.95.024426}.

Less exotic systems with known groundstates may also be interesting to study. 
For instance, the phase diagram of the $J_1-J_2$ antiferromagnetic Heisenberg chain could be investigated, where $J_1$ and $J_2$ are the nearest and next-nearest neighbour exchange couplings, respectively.
When $J_2/J_1$ is small, the chain realizes a Tomonaga-Luttinger liquid. 
However, as $J_2/J_1$ is increased, the system becomes more frustrated.
At sufficiently large $J_2/J_1$, the chain undergoes a spin Peierls transition to a valence bond solid \cite{Haldane82,sachdevBook}. 
These two phases can be distinguished by their excitation spectrum: while the Tomonaga-Luttinger liquid has gapless excitation, the valence bond solid has a gapped spectrum.
As a one dimensional model, the $J_1-J_2$ model may be experimentally accessible on a shorter timescale. 

Another possibility is to simulate the Heisenberg model on either the honeycomb or square lattices.
Both models are known to possess N\'{e}el order at zero temperature [see Figs.~\ref{fig:trivalentModels}(d) and~\ref{fig:4bond}(b) for possible implementations].
After allowing these or similar systems to attain their groundstates, the spin moment on every site could be independently ascertained, allowing on-site resolution of the alternating order parameter in a manner analogous to the cold atoms experiments of Ref.~\citenum{Mazurenko17}.

\subsubsection{Disordered magnets}\label{sec:disorderMag}

Since it is present in all matter to some degree, disorder is a far-reaching topic.
Nevertheless, most studies of many-body quantum states entirely neglect its effects, in part because its inclusion often renders the problem intractable.
Similar to the clean limit just discussed, computational limitations restrict our ability to numerically simulate many of the systems of greatest interest.

The great tunability of the system parameters makes our quantum simulator an ideal platform to explore the effects of disorder.
Notably, the implementation of random bond disorder of various forms differs little from the implementation of the pure systems mentioned above.
Within our setup, both the disorder strength and symmetry can act as tuning parameters.
Variations in the exchange coupling strength can be tuned both through tunneling amplitudes and the charging energy, with the requirement that $t\ll E_C$ as a sole constraint.
Depending on the model under study, random changes of the sign of the exchange coupling $J$ may be attainable; however, given $J$'s dependence on the geometry of the architecture, this may be more difficult to achieve than a change in magnitude (see Sec.~\ref{sec:ExCouplingSign} for details).
An additional advantage of the Majorana simulator is that, unlike cold atoms systems \cite{Schreiber15,Choi16}, both true randomness and quasi-periodicity can be achieved (we comment on this further below).
Weak random-field disorder could also be introduced either by allowing direct (and very weak) tunneling between Majoranas on a single island [see Eq.~\eqref{eqn:EffZeeman}] or by the more sophisticated methods using quantum dots discussed in Sec.~\ref{sec:add-quantum-dots}.

Disordered phenomena in one dimensional systems is relatively well understood.
In particular, there exists a large body of work on random quantum Ising \cite{Fisher92,Fisher95} and Heisenberg spin-1/2 chains \cite{Fisher94,Hirsch80}.
An especially interesting phase that is should appear is the ``random singlet phase" \cite{PhysRevB.22.1305,PhysRevLett.48.344,Fisher94}, which is characterized by the existence of weak singlets between arbitrarily distant spins.
Features of this phase such as the quadratic decay of correlation functions, $\sim r^{-2}$, \cite{Fisher94} may be possible to observe in our setup.

The effects of disorder in frustrated magnets is far more complex in higher dimensions, and this is precisely the regime where our simulator may be most useful. 
For example, the authors of Ref.~\citenum{kimchi17} conjectured a Lieb-Schultz-Mattis-like theorem \cite{LIEB1961407,PhysRevB.69.104431,PhysRevLett.78.1984,PhysRevLett.84.1535} for disordered systems: when translational symmetry is preserved on average and the average unit cell contains a single spin-1/2, a disordered paramagnet must either possess gapless spin excitations or be topologically ordered.
Either analytically or numerically demonstrating this hypothesis is very challenging, and a sufficiently large Majorana island network could provide new insight.

Disordered systems may also be problematic from an experimental standpoint.
Determining the cause of power law behaviour in observables such as the susceptibility can be difficult, and this can make distinguishing a disordered state from a quantum-critical one challenging \cite{Singh10,CastroNeto98}.
For  systems  like  Herbertsmithite (a material believed to be described by the
kagome  Heisenberg  model) this is particularly disadvantageous since, as discussed, the groundstate of this Hamiltonian in the clean limit may be a Dirac spin liquid, a state described by a conformal field theory.
Numerical simulations of the kagome Heisenberg model with random bond disorder \cite{kawamura14} appear to indicate that sufficient disorder suppresses magnetic order, favouring a 2$d$ version of the random singlet phase.
The authors of Ref.~\citenum{kawamura14} claim that their results are in agreement with the dynamic structure factor obtained via INS \cite{NormanKagome16,kagomeINS_Helton10}.
This is counter to the reasoning of Han \emph{et al.} \cite{HanNorman16}, who assert that the  signal is instead the consequence of separate contributions from the $\Zt$ spin liquid and the impurities within the sample. 
The type of large-scale simulations our Majorana network may provide could help resolve this issue [see Fig.~\ref{fig:4bond}(c)].
Further, different types of disorder can also be studied and possibly matched onto the experimental data.

\subsubsection{Quantum criticality}

Another application of our setup is to systems undergoing quantum phase transitions.
The traditional first example of a phase transition is the critical point separating the ferromagnetic and paramagnetic phases of the Ising model. 
The 1+1$d$ version is perhaps the simplest system our network can realize since it can be simulated using the the simplified architecture discussed in Sec.~\ref{sec:SimpArch} and shown in Fig.~\ref{fig:flat_Kitaev}.

XXZ spin chains are another example of an interesting, yet comparatively simple realization of quantum criticality. 
Their critical points are described by a set of conformal field theories known as the SU(2)$_k$ Wess-Zumino-Witten nonlinear $\s$-models.
The chain in Fig.~\ref{fig:trivalentModels}(e) would correspond to $k=1$, and larger $k$ conformal field theories are obtained by considering larger effective spins \cite{Affleck86,Affleck88,henkel2013}.

Phase transitions can also be tuned by introducing anisotropic bonds. 
The square lattice Heisenberg model with exchange coupling $\cJ$ on the horizontal bonds of every other column and exchange coupling $g\cJ$ on all remaining bonds is one example \cite{sachdevReview09,anisoSqLat1,anisoSqLat2}.
In the limit $g\to0$, a valence bond solid is realized, whereas in the isotropic limit, $g\to1$, the groundstate has N\'{e}el order. 
At intermediate $g$ a quantum phase transition in the same university class as the classical 3$d$ Heisenberg model is realized \cite{anisoGen1}.
Similarly, continuous transitions belonging to different, less well-studied, universality classes can be induced on frustrated lattices such as the Shastry-Sutherland lattice \cite{anisoGen1,anisoGen2}. 

Our setup is also capable of capturing more exotic phase transitions, whose descriptions require emergent gauge fields.
These deconfined critial points \cite{Senthil2004a,Senthil2004b} go beyond the Landau-Ginzburg paradigm and are not necessarily associated with the breaking of a symmetry.
They can describe critical points connecting different symmetry-broken states as well as transitions involving topologically ordered phases. 
A potential candidate is the Heisenberg-Kitaev honeycomb model mentioned above: numerics indicate that a continuous and potentially exotic phase transition could lie between the spin liquid and stripy phases \cite{KitaevHeisen_Schaffer2012,KHmodel_Chaloupka10,KHmod_Reuther11}.

Disordered critical systems can also be studied, and the disorder strength may in fact be treated as another tuning parameter. 
This is often a very difficult problem since disorder tends to be relevant under the renormalization group flow, taking the system outside of the perturbative regime \cite{sachdevBook,AharonyDisorder18a,AharonyDisorder18b}.
Even in the 2+1$d$ Ising model, the fate of the system with random bond disorder is not well-understood \cite{sachdevBook}.
It may also be possible to obtain an interacting and disordered fixed point proximate to the Dirac spin liquid phase by adding disorder to the kagome Heisenberg model \cite{qedDisorder_thomson17,qedDisorder_Raghu}.
The form of disorder necessary to obtain this critical theory is specific enough that it is unlikely to be realized in Herbertsmithite, and so the high degree of control afforded by the Majorana network makes it an ideal setting for the observation of this type of critical theory.

\subsubsection{Entanglement entropy}\label{sec:EntEntropy}

The most intriguing aspect of quantum systems is their nonlocal nature, and a widely-used measure of this is the entanglement entropy.

Obtaining non-local state information is incredibly difficult experimentally.
Nevertheless, recent cold atom experiments have managed precisely this \cite{Islam15,Kaufman16,Daley12}, indicating that it should be possible in our Majorana network as well.
In fact, the protocol used by the authors of Refs.~\citenum{Islam15} and~\citenum{Kaufman16} to determine the second Renyi entropy should also be applicable to a Majorana network simulating spin-1/2 chains. 
As in their experiment, one would arrange two spin chains like in Fig.~\ref{fig:trivalentModels}(e) alongside each other and mimic the ``beamsplitting" procedure by coupling them in the appropriate manner for a certain duration. 
As we have stressed, an advantage our platform has over optical lattice experiments is the ability to tune the interactions.
This offers the appealing possibility of measuring the entropy for a larger variety of spin models.

\subsection{Dynamical properties}\label{sec:DynProp}

Our discussion has so far focused on systems at equilibrium.
However, many of the most important recent advances relate to out-of-equilibrium behaviour.
Since the tunneling amplitudes between islands are controllable, the Hamiltonian may also be varied in time.
While we have already mentioned exploiting this to obtain equilibrium-state information via linear response [see Sec.~\ref{sec:CorrFunMeas}], our system is not restricted to this regime, and non-perturbative effects should also be accessible.

A common way to characterize these dynamics is by studying the system's evolution after quickly altering the Hamiltonian.
When the change is abrupt and persists for all subsequent times, it is typically called a ``quench."
Conversely, the term ``pulse" is used to describe a change with a small duration.
Depending on the situation, these quenches and/or pulses could be either local or global.
These parameter changes can be implemented reasonably quickly ($\sim 20-30$ ps) \cite{Nakamura1999} relative to the fastest timescale of the effective spin Hamiltonian ($\tau_J=h/J\sim 60$ ps).
The isolation of the system is limited by the quasiparticle poisoning time, $\tau_\mathrm{qpp}\sim 1$ ms.
Since this is orders of magnitude larger than the tunneling time, $\tau_\mathrm{qpp}/\tau_J\sim 10^7$, the system has more than enough time to evolve in \emph{de facto} isolation for novel physics to be uncovered.

In contrast to an experiment being performed for groundstate determination, it is important to know the initial state before the quench or pulse is performed in a simulation probing nonequilibrium behaviour.
One simple possibility is to have each vertex of the network initialized to a definite spin value resulting in a total product state.
More complicated initial states can be obtained by implementing the procedure outlined in the previous section: start with an arbitrary state and turn on a Hamiltonian with the desired groundstate. 
Once that groundstate is achieved, the quench and/or pulse can be implemented.

We now discuss two broad topics of study: the scrambling of information and many-body localized phases.

\subsubsection{Scrambling and quantum chaos}

Recently, the process of thermalization has received renewed interest: given unitary time evolution, how does dissipation emerge? 
The eigenstate thermalizaton hypothesis (ETH) states that an isolated system initially far from equilibrium (for instance, after a quench), approaches a state in which observables appear thermal \cite{srednickiETH,dAlessio15}.
The approach to thermal equilibrium is very intimately related to spreading of entanglement and the scrambling of information. 

We discussed in the previous section how the protocol used to measure the entanglement entropy in the cold-atom experiments of Ref.~\citenum{Islam15} and~\citenum{Kaufman16} could be applied to our setup to measure the entanglement of certain spin chains.
In addition, the experiment of Ref.~\citenum{Kaufman16} not only measured the entanglement entropy, but its evolution.
They applied a quench to a 1$d$ row of atoms and measured the second Renyi entropy associated with half of their system as a function of time.
They observed an increase in subsystem entropy and attributed this to the entanglement between subsystems. 
In accordance with the ETH, the entanglement entropy approached the thermal entropy.
These experiments and observations should be reproducible with the Majorana network simulator, and the additional control furnished by the Majorana device should allow similar studies to be done in more diverse settings.
For instance, the direct observation of some of the analytical results obtained for 1+1$d$ conformal field theories may be accessible \cite{Calabrese05,Calabrese09,Calabrese16,Asplund15}.

Out-of-time-order correlators (OTOC) have recently become an especially powerful way to obtain information on the form and evolution of scrambling.
The OTOC can be thought of as a measure of the failure of two operators initially at distinct points (say, $x$ and $y$) to commute at a later time $t$: $\Braket{[A_x(t),B_y(t)]^\dag [A_x(t),B_y(t)]}$, where $A_x(0)$ ($B_y(0)$) is localized at $x$ ($y$).
The OTOC is a difficult object to obtain experimentally since it appears to require an reverse time-evolution step. 
Nevertheless, there are a number of proposals for measuring the OTOC.
Some explicitly require that the Hamiltonian be taken to minus itself, usually through a spin echo protocol \cite{Zhu16,Swingle16}.
Analogous measurements could be implemented in our system through the techniques of Refs.~\citenum{Petta05} and~\citenum{Nadj-Perge2010}.
Another intriguing possibility was proposed in Ref.~\citenum{Yao16} where the authors suggested measuring the OTOC through the presence of an ancillary spin within  the setup for the entanglement measurement experiments mentioned above \cite{Islam15,Kaufman16}.

\subsubsection{Many-body localization and prethermalization}\label{sec:MBL}

In contrast to the systems mentioned in the previous section, many-body localization (MBL) occurs when systems do \emph{not} thermalize, even at long times \cite{alet18,Altman15,Huse15}.
In contrast, a classical glass is a phase of matter that thermalizes exponentially slowly.
Since the quasi-particle poisoning time $\tau_\mathrm{qpp}$ can be seven orders of magnitude larger than the tunneling timescale $\tau_J= h/J$, our simulator should be capable of distinguishing the MBL phase from a classical glass.

Some of the simpler implementations of our proposal are especially well-suited to studying the MBL phase.
A Heisenberg spin chain with nearest-neighbour couplings and random local Zeeman fields is the paradigmatic example of a system able to support MBL \cite{alet18,Huse15}.  
The local fields can be realized by simply adding tunnel couplings between Majoranas on the same island or using the quantum dot protocols of Sec.~\ref{sec:ExpRealization}.
Importantly, as already stressed in Sec.~\ref{sec:disorderMag}, choosing an arbitrary distribution of fields is straightforward.

Possible signatures of a many-body localized phase in a spin chain can be observed in the temporal decay of an initially polarized spin state after a quench. 
In our case, the single-site spin polarization can be inferred from local parity measurements on a Majorana island. 
A possible experiment could measure, say, $\braket{\sigma^z_i(t)}$ after the tunneling amplitudes linking site $i$ to the rest of the system have been switched on for a duration $t$. 
The time dependence of the disorder-averaged expectation value serves as a measure of localization \cite{Agarwal2015}. Such experiments are essentially equivalent to decoherence measurements of qubits, which have been demonstrated in semiconducting quantum dots \cite{Petta05}.

Alternatively one could detect global quantities such as the staggered magnetization of a system initialized in a N\'eel state (this is similar to the suggestion at the end of Sec.~\ref{sec:FrustMag}). In the MBL phase, such global observables are expected to retain a nonzero value after disorder averaging, even in the long-time limit. 
Such experiments would be similar to the cold-atom experiments of Refs.~\citenum{Schreiber15} and~\citenum{Choi16}, which measured the decay of a density polarization of interacting fermions over time. 
Notably, MBL physics was observable after only $30-50\tau_J$, well within reach of our simulator (our conventions differ slight from those of Refs.~\citenum{Schreiber15} and~\citenum{Choi16}, where they define the tunneling time as $\tilde{\tau}_J= 2\pi\tau_J$). 

Unlike a thermal system, the entanglement entropy of an MBL phase obeys an area law  \cite{alet18,Altman15,Huse15}.
Once more, by mimicking the scheme of Refs.~\citenum{Islam15} and~\citenum{Kaufman16}, this should be observable with our Majorana simulator.
More interesting the expected logarithmic growth of entanglement after a quantum quench could also be seen \cite{Bardarson12}.
Since this behaviour is only apparent at exponentially long times, the long coherence time of the Majorana network is crucial.

Another intriguing question is to what extent localization is able to protect symmetry-protected topological order at finite temperatures \cite{Chandran2014}.

Our ability to tune disorder also allows for systematic studies of the many-body localization transition.
As it separates a thermal phase from  the non-thermal MBL phase, standard statistical methods are not applicable. 
For instance, the transition should only be visible using dynamic probes, which is why it is typically referred to as a ``dynamical phase transition." 
One predicted feature of this transition is an exponentially slow thermalization time in the critical region \cite{Parameswaran17} making long coherence times especially crucial in this context.
With $\tau_\mathrm{qpp}\sim 10^7 \tau_J$, our simulator should be able to probe this and other novel features of the transition.
Further, our ability to control not only the strength of the disorder, but also its form suggests the intriguing possibility of tuning between quasi-periodic and random distributions, which have transition believed to belong to different universality classes \cite{Khemani17}.

Spin chains systems are also a particularly suitable platform to study prethermalization. This phenomenon describes systems which remain in a metastable nonequilibrium state for exponentially long times until they ultimately thermalize. Our quantum simulator combines two essential prerequisites for observing prethermalization: tunability and coherence. Good control over experimental parameters is needed to customize the exchange interactions, allowing  the system to enter the prethermal regime. In addition, observing exponentially long time scales requires long coherence times of the spin.
An intriguing signature of prethermalization within the scope of our simulator is the stability of edge spins in the transverse-field Ising model \cite{Else2017}.

\section{discussion}\label{sec:discussion}

In this paper, we have proposed a quantum simulator of spin systems whose spins are topologically protected from magnetic field noise. 
We began with a description of how to simulate a large variety of spin models using Majorana islands.
We subsequently proceeded to outline concrete experimental platforms based on existing technology.
We followed by proposing detailed protocols for the characterization of the simulated model within the bounds of current experimental capabilities.
Finally, we discussed a number of important applications for which our simulator may further our understanding of many-body physics.
We conclude our paper with a comparison to other quantum simulators in the literature and an outlook to the future. 

\subsection{Comparison to other quantum simulators}
While constructing the Majorana network we describe may be challenging, we emphasize that all the difficulties inherent in our proposal must eventually be overcome in the construction of a topological quantum computer.
More than anything, our quantum simulator stands as a stepping stone on the path to universal topological quantum computation.
We have shown that with even relatively few Majorana qubits, new physics can be uncovered. 

The analogue quantum simulator we have proposed generically requires fewer resources than its digital counterparts where Hamiltonian time evolution must be Trotterized through the application of a  series of quantum gates on given initial states. 

In fact, Kassal \etal \cite{Kassal11} have shown that at least 100 qubits and 200,000 operations per time step are required to simulate a Hamiltonian with pairwise interactions better than a classical computer.
While no analogue simulator can compete with the versatility this type of simulator imparts, these requirements are well outside of current experimental capacity. 

Relative to other analogue simulators, the close relation between the Majorana simulator and a topological quantum computer grants our proposal several advantages.
In particular, we address the extremely long coherence times, the fine-tunability, the low temperatures, and the straightforward relaxation mechanisms our simulator provides.

We have repeatedly stressed the length of the decoherence time (set by quasi-particle poisoning) relative to the timescale set by a typical exchange constant: $\tau_\mathrm{qpp}/\tau_J\sim 10^7$.
In comparison, for the study of MBL, the cold-atom experiments were able to attain coherence times of approximately $\sim30-50\tau_J$ \cite{Schreiber15,Choi16}.
While features of the MBL phase were visible in these experiments, definitively distinguishing the MBL phase from a classical glass necessitates coherent evolution for time scales exceeding $\tau_J$ by several orders of magnitude, something the Majorana simulator provides. 
Further, with access to such long time scales, a number of experiments probing the precise nature of the phase and its transition should be possible. 
In particular, in Sec.~\ref{sec:MBL}, we mentioned being able to observe the exponentially slow thermalization time in the critical region of the MBL transition \cite{Parameswaran17} as well as the the logarithmic entanglement growth of the MBL phase \cite{Bardarson12}.
Conversely, given the current decoherence times, neither experiment is feasible with cold atoms \cite{quSimUltraCold}. 

The high-degree of control our platform offers is another advantage.
In the language of hardware design, the simulator we propose follows the ``bottom-up" philosophy. 
In contrast, some of the most successful simulators take the opposite, ``top-down" approach.
Experiments with cold atoms in optical lattices and trapped ions belong to the latter category since they let the particles interact naturally.
As a result, this type of simulator possesses fewer tuning parameters, and this makes breaking the SU(2) spin symmetry in these systems much more difficult. 
It is unlikely that a model like the Heisenberg-Kitaev model [see Eq.~\eqref{eqn:HKham}] could be engineered in an optical lattice.
Similarly, true randomness is difficult to achieve in optical lattices, and quasi-periodicity is favored instead \cite{Schreiber15,Choi16}.

The Majorana network is also able to access low temperatures.
For instance, while ultracold Fermi gases have observed N\'{e}el order in the square lattice Hubbard model \cite{Mazurenko17}, attaining a sufficiently low temperature was very challenging.
The lowest they were able achieve was $T/J\cong0.45$, where $J$ is the effective exchange coupling.
While adequate for the observation of antiferromagnetism on the square lattice, this temperature is not low enough to resolve important features of certain spin liquids.
As an example, in Sec.~\ref{sec:FrustMag}, we mentioned that the groundstate of the kagome Heisenberg model is currently unknown and that the spin gap was a clear distinguishing feature of the two main candidate states.
However, should it exist, the estimated gap is very small, with numerics currently setting it around $0.05-0.1J$ \cite{mendels11,kagomeDMRGDependbrock,Nishimoto13}.
This would not be visible at the temperatures currently accessible in Fermionic quantum gases.
Conversely, the expected temperature of our proposal is an order of magnitude smaller, $T/J\sim 0.05-0.07$, making the observation of the debated spin gap a reasonable objective.

Finally, unlike many quantum simulators, the platform we propose does not require complex mappings between the degrees of freedom of the physical parameters and the model of interest.
Since the environment may interact with the simulator in fundamentally different ways than the original model, this can result in complications when attempting to relax to the groundstate \cite{Brown07}.
For instance, trapped ions systems can simulate spin systems, but only after transforming to the interacting picture  \cite{Kim10,Porras04,Deng05,Friedenauer08} where the decoherence process is very different than in a physical spin system. 

A number of other proposals to use Majoranas as simulators have been made. 
In particular, at first glance the Majorana network proposal of Ref.~\citenum{Barkeshli15} may appear very similar.
However, their scheme does not enjoy the same protection against decoherence that ours does.
More importantly, in order to obtain the two-fold degenerate subspace needed to form an effective spin, the backgate voltage of each island must be precisely tuned.  
We use a similar setup in Sec.~\ref{sec:split-island}, but only for the purpose of measurement.
Similarly, the proposal of 
Ref.~\citenum{Kells2011} constructs a Majorana network described by an effective quadratic Hamiltonian.
As per Kitaev's solution, such a Hamiltonian can only simulate a \emph{single} flux sector of the full model, and, as such, their proposal is unable to reproduce the full dynamics of the Kitaev models.

The realization of spins with Majorana islands has been exploited in several recent works as well. 
The initial discussion focused on the topological Kondo effect in a Majorana island \cite{Beri2012}. 
More recent works have proposed architectures for the realization of logical qubits in surface \cite{Vijay2015,Landau2016} or color codes \cite{Litinski2017}.
While these development are of prime importance for the ultimate construction of a topological quantum computer, the hardware requirements are very demanding.
For instance, the authors of Ref.~\citenum{Litinski2017} estimate that their colour code implementation requires 500 physics logical qubits for every physical qubit. Our proposal, in contrast, has the potential to solve important problems with far fewer qubits.

\subsection{Future directions}

We have demonstrated that a multitude of interesting spin models can be assembled from a basic setup comprised of superconducting islands with four Majoranas each.
Nevertheless, models with next-nearest neighbor interactions are typically difficult to construct, requiring a large number of islands per site. 
It would be interesting to study how different superconducting island constructions (\emph{e.g.}, islands with a larger number of Majoranas) could be used to extend the scope of our platform.
As an example, a setup containing six Majoranas per island was recently proposed in an independent work \cite{Sagi2018} as a realization of the Yao-Kivelson model.

Devising ways to engineer additional, further-neighbour couplings between spins is another meaningful direction of study. 
An interesting experimental question is whether out-of-plane couplings between Majoranas can be fabricated.
This would be useful for both models in three dimensions as well as two-dimensional models with further-neighbour interactions like the $J_1-J_2$ Heisenberg model on the square lattice.
One could also consider adding long-range interactions between spins by coupling multiple islands to the same delocalized mode, such as a photonic mode in a resonator or electrons in an extended lead.

The ability to couple effective spins to new degrees of freedom  may also be interesting in its own right. 
For example, the Dicke model, the paradigmatic model for a superradiant quantum phase transition, could be realized by allowing several Majoranas to interact with a single transmission line resonator \cite{Garraway2011}.
Moreover, our quantum simulator is capable of reproducing the Ising-Dicke model, for which a rich phase diagram results from the competition between antiferromagnetic exchange and coupling to the oscillator mode \cite{Gammelmark2011,Zhang2014}. 
Decoherence in this case is presumably limited to losses in the resonator rather than decoherence of the spins.

With some adjustments, our simulator may also be capable of reproducing Kondo lattice physics.
In one dimension, this could be implemented by contacting the Majoranas to three leads \cite{Beri2012} such that all Majoranas with the same label interact with the same lead. Kondo physics should be obtained by lowering the electron density of the leads to match the density of effective spins.
While the situation is more complicated in two dimensions, one could perhaps couple the Majorana islands to an array of quantum dots \cite{Manousakis02,Byrnes07,Singha11}.
By arranging the dots so that each Majorana is associated to a cluster of three dots labelled $x$, $y$, and $z$, a two-dimensional Kondo lattice model should be obtained upon allowing each dot to interact with the associated $b$-Majoranas and the nearest-neighbour dots of the same kind.

The experimental capability to generate ac drives discussed in  Sec.~\ref{sec:ExperiProbes} points to Floquet systems as another intriguing direction of study.
By applying a continuous drive to our system, novel nonequilibrium states of matter and dynamical phase transitions should be accessible. 
The Majorana simulator seems particularly suitable for the study of Floquet phases as the long coherence time of the effective spins is expected to reduce the heating effects of the drive. 
Moreover, the ability to tune the degree to which the system couples to the environment might be useful for studies of driven-dissipative systems. 

The observation of multipoint correlators may be an interesting extension to the proposed measurement protocols.
Such measurements could be based on the pump-probe technique proposed in Sec.~\ref{sec:CorrFunMeas}, in which physical observables were obtain through the detection of the ac response of a spin expectation value to a locally applied alternating Zeeman field. 
In addition to signals at the drive frequency, one can look for signals at higher harmonics. 
Alternatively, one could apply two incommensurate drives at different locations and detect a signal at the sum or difference of the applied frequencies. 
Such measurements should be sensitive to multipoint correlation functions.

In Sec.~\ref{sec:EntEntropy} we suggested that the second Renyi entropy of a spin chain could be obtained using a  protocol similar to the what was done in the cold atom experiments of Refs.~\citenum{Islam15} and~\citenum{Kaufman16}. 
It is natural to ask if other Renyi entropies could be observed and whether this scheme can be adapted to 2$d$ lattices. 
Similarly, we argued that proposals designed to measure the OTOC in cold atom systems were also applicable to our setup.
It would be interesting to identify experimental strategies that take advantage of the semiconductor realization of our simulator, such as through nonlocal transport signatures.

\section*{Acknowledgments}

We are grateful to Shubhayu Chatterjee, Bertrand Halperin, Chetan Nayak, Florentin Reiter, Subir Sachdev, and Seth Whitsitt for valuable discussions. F.~P.\ acknowledges financial support by the STC Center for Integrated Quantum Materials, NSF Grant No.\ DMR-1231319.
A.~T. acknowledges support by the NSF under Grant No.\ DMR-1664842.

{\em Note added.}\ --- During the completion of the manuscript, we became aware of Ref.~\onlinecite{Sagi2018}, which discusses a distinctly different array Majorana islands that realizes the Yao-Kivelson model and proposes a measurement scheme that bears some overlap with our work.

\appendix

\begin{widetext}
{
\setstretch{1.3}

\section{Numerical solution to 4-bond vertex}\label{app:4bondSims}

In this appendix we provide details of the numerical calculation used in Sec.~\ref{sec:4bond}. 
We begin by discussing the Hilbert space and Hamiltonian before presenting a broader  study of the relevant parameter regime.

The charge degree of freedom will be given by $\v{N}=(N_1,N_2,N_3)$ where $N_i$ is the difference between the number of electrons on each island and the integer closest to the charge induced by the backgate voltage.
That is, we write $Q_{0,i}/e=N_i+\d N_i$ where $N_i\in\mathds{Z}$ and assume that $\abs{\d N_i}\ll 1$.
Then we can express the action of the charging Hamiltonian as
\eq{
H_{C,i}\Ket{N_i}=E_{C,i}\(\hat{n}_{e,i}-{Q_{0,i}\o e}\)^2\Ket{N_i}=E_{C,i}\( N_i - \d N_i\)^2 \ket{N_i}
}
where $E_{C,i}={e^2/ 2C_i}$.
Since $\abs{\d N_i}$ is small, the ground state manifold of $H_C$ for the three island  has $\vN=(0,0,0).$

We describe the Majorana degrees of freedom in the basis defined by the complex fermions
\eq{
f_{iA}&={1\o2}\(c_i+ib_i^x\),
&
f_{iB}&={1\o2}\(b_i^z+ib_i^y\).
}
The state of the island also depends occupation of $\vn_i=(n_{iA},n_{iB})$ on each island, $i=1,2,3$.
The Majoranas act on this basis as
{\small
\eq{
b_1^x\Ket{\vn_1;\vn_2;\vn_3}&=i(-)^{n_{1A}}\Ket{1-n_{1A},n_{1B};\vn_2;\vn_3},
&
b_1^y\Ket{\vn_1;\vn_2;\vn_3}&=i(-)^{\vn_{1}}\Ket{n_{1A};1-n_{1B};\vn_2;\vn_3},
\nt
c_1\Ket{\vn_1;\vn_2;\vn_3}&=\Ket{1-n_{1A},n_{1B};\vn_2;\vn_3},
&
b_1^z\Ket{\vn_1;\vn_2;\vn_3}&=(-)^{n_{iA}}\Ket{n_{1A};1-n_{1B};\vn_2;\vn_3},
\nt
b_2^x\Ket{\vn_1;\vn_2;\vn_3}&=i(-)^{\vn_{1}+n_{2A}}\Ket{\vn_1;1-n_{2A},n_{2B};\vn_3},
&
b_2^y\Ket{\vn_1;\vn_2;\vn_3}&=i(-)^{\vn_{1}+\vn_2}\Ket{\vn_1;n_{2A};1-n_{2B};\vn_3},
\nt
c_2\Ket{\vn_1;\vn_2;\vn_3}&=(-)^{\vn_1}\Ket{\vn_1;1-n_{2A},n_{2B};\vn_3},
&
b_2^z\Ket{\vn_1;\vn_2;\vn_3}&=(-)^{\vn_1+n_{2A}}\Ket{\vn_1;n_{2A};1-n_{2B};\vn_3},
\nt
b_3^x\Ket{\vn_1;\vn_2;\vn_3}&=i(-)^{\vn_{1}+\vn_2+n_{3A}}\Ket{\vn_1;\vn_2;1-n_{3A},n_{3B}},
&
b_3^y\Ket{\vn_1;\vn_2;\vn_3}&=i(-)^{\vn_{1}+\vn_2+\vn_3}\Ket{\vn_1;\vn_2;n_{3A};1-n_{3B}},
\nt
c_3\Ket{\vn_1;\vn_2;\vn_3}&=(-)^{\vn_1+\vn_2}\Ket{\vn_1;\vn_2;1-n_{3A},n_{3B}},
&
b_3^z\Ket{\vn_1;\vn_2;\vn_3}&=(-)^{\vn_1+\vn_2+n_{3A}}\Ket{\vn_1;\vn_2;n_{3A};1-n_{3B}},
}}

Since these occupation numbers correspond to physical excitations, they are naturally dependent on the charge degrees of freedom.
We assume that all excitations on the islands conserve charge (no quasi-particle poisoning).
Without loss of generality, we choose to work in a basis where islands with an even charge have an even number of occupied Majorana modes.
That is, $\vn_i$ is restricted to $(0,0)$ and $(1,1)$ ($(0,1)$ and $(1,0)$) when $N_i$ is even (odd).
This constraint is, of course, dependent on our definition of $f_{iA}$ and $f_{iB}$.
Each value of $\vN$ labels an eight-dimensional subspace parametrized by the Majorana occupation numbers.
Keeping the dependence of the $\vn_i$'s in mind, we write the states as $\Ket{\vn_1;\vn_2;\vn_3}\Ket{\vN}$. 


The total Hamiltonian we simulated is
\eq{\label{eqn:Hsim}
H_{\mathrm{fluc}}&=\sum_{i=1}^3H_{C,i}+H_{\mathrm{tun},12}+H_{\mathrm{tun},23}+H_J.
}
[This is the same Hamiltonian as given in Eq.~\eqref{eqn:ExactVertexHam}.]
The tunneling Hamiltonian given in Eq.~\eqref{eqn:tunHam} acts on these states as
\eq{\label{eqn:HtunBasis}
H_{\mathrm{tun},12}&\Ket{\vn_1;\vn_2;\vn_3}\Ket{\vN}
={it\o2}\Bigg[(-)^{n_{1B}}\Big( (-)^{n_{1A}}-(-)^{n_{2A}} \Big)\Ket{1-n_{1A},n_{1B};1-n_{2A},n_{2B};\vn_3}
\nt
&\quad
-(-)^{n_{2A}+n_{2B}}\Ket{n_{1A},1-n_{1B};n_{2A},1-n_{2B};\vn_3}\Bigg]\otimes
\Big[\Ket{N_1+1;N_2-1;N_3}
+\Ket{N_1-1;N_2+1;N_3}\Big],
\nt
H_{\mathrm{tun},23}&\Ket{\vn_1;\vn_2;\vn_3}\Ket{\vN}
={it\o2}\Bigg[(-)^{n_{2B}}\Big( (-)^{n_{2A}}-(-)^{n_{3A}} \Big)\Ket{\vn_1;1-n_{2A},n_{2B};1-n_{3A},n_{3B}}
\nt
&\quad
-(-)^{n_{3A}+n_{3B}}\Ket{\vn_1;n_{2A},1-n_{2B};n_{3A},1-n_{3B}}\Bigg]\otimes
\Big[\Ket{N_1;N_2+1;N_3-1}
+\Ket{N_1;N_2-1;N_3+1}\Big],
}
where we've used the fact that the phase operator $e^{i\phi_i/2}$ adds a charge to the $i$th island, e.g. $e^{i\phi_1/2}\Ket{N_1,N_2,N_3}=\Ket{N_1+1,N_2,N_3}$.
For notational simplicity, we've omitted the ``vert" subscript when writing the tunneling amplitude, simply using $t$.

For sufficiently small tunnel couplings, the Josephson energy can be neglected. 
However, for large values of $t$ it can have an appreciable effect and should be considered:
\eq{
H_J&=-E_J\big[ \cos\(\phi_1-\phi_2\) + \cos\(\phi_2-\phi_3\)\big].
}
$H_J$ acts on the charge degree of freedom as
\eq{
H_J\Ket{\vN}&=-{E_J\o2}\bigg[
\Ket{N_1+2,N_2-2,N_3}+\Ket{N_1-2,N_2+2,N_3}
\nt&\quad
+\Ket{N_1,N_2+2,N_3-2}+\Ket{N_1,N_2-2,N_3+2}
\bigg].
}
The Josephson energy, $E_J$, is expected to depend linearly on the transparency $T$ of the system while the Majorana tunneling, $t$, should have a linear dependence: $t\propto T$, $E_J\propto \sqrt{T}$.
We will therefore express $E_J$ in terms of $t$ as
\eq{
E_J=\ep_J t^2,
}
for some proportionality constant $\ep_J$.

Finally, in addition to the capacitance of the single islands, they will have a mutual capacitance:
\eq{
H_{C,\mathrm{mut}}&=E_{C,\mathrm{mut}}\(\hat{n}_{e,2}-{Q_{0,2}\o e}\)\Bigg[
\(\hat{n}_{e,1}-{Q_{0,1}\o e}\)+\(\hat{n}_{e,3}-{Q_{0,3}\o e}\)
\Bigg],
}
which acts on $\Ket{\vN}$ as
\eq{
H_{C,\mathrm{mut}}\Ket{\vN}&=E_{C,\mathrm{mut}}\(N_2-\d N_2\)\bigg[ \(N_1-\d N_1\)+\(N_3-\d N_3\)\bigg]\Ket{\vN}.
}
We find that this term has little effect, and for this reason it is not included in Eq.~\eqref{eqn:Hsim}.

There are several points to be made.
First, our simulations have to be truncated at some $N_\mathrm{max}=\max\abs{N_i}$. 
That is, we only consider states where $\abs{N_i}\leq N_\mathrm{max}$. 
Also important is that in all parameter regimes and all values of $N_\mathrm{max}$ we consider, we find that $H_\mathrm{fluc}$ has a two-fold degenerate groundstate. 

The appropriate choice of $N_\mathrm{max}$ depends on the parameters of the Hamiltonian.
When $E_C$ dominates the problem, there will be few fluctuations in $\vN$, allowing $N_\mathrm{max}$ to be relatively small, maybe even 3 or 2.
However, the converse situation is also important to consider. 
In $H_\mathrm{fluc}$ this is simply achieved by ``setting" $E_C=E_J=0$. 
More correctly, this should be thought of as the limit $E_J\to\infty$ since when only the tunneling Hamiltonian is present, there ceases to be a distinction between islands~1-3, and the system should instead be viewed as a single large island.
This pins the superconducting phases to a common value, and we can set $\phi_1=\phi_2=\phi_3=0$ without loss of generality.
Because total charge is still conserved, the value of $E_C$ ceases to matter. 

The role of $H_\mathrm{tun}=H_{\mathrm{tun},12}+H_{\mathrm{tun},23}$ [Eq.~$\eqref{eqn:HtunBasis}$] is to couple the $c_i$'s to one another, $b_i^x$'s to one another, and the $b_i^y$'s to one another.
It follows that in the groundstate we can identify $C\sim c_1\sim c_2\sim c_3$, $B^x\sim b_2^x\sim b_2^x \sim b_3^x$, and $B^y\sim b_1^y\sim b_2^y\sim b_3^y$.  
Unlike the rought argument given in Sec.~\ref{sec:4bond}, without the charge constraint, there is no reason to treat the $b_i^z$'s as a single degree of freedom.
As a result, we have six non-interacting Majorana modes with a single charge constraint ($C$, $B^x$, $B^y$, and the three $b_i^z$'s), implying that the groundstate should be $2^{3-1}=4$-fold degenerate.
This four-fold degeneracy is found exactly when we diagonalize $H_\mathrm{tun}$ in the basis which treats only the Majorana degrees of freedom.
In particular, numerically diagonalizing 
\eq{
H_{\mathrm{tun},NC}=(t+\bar{t})\bigg[ ib_1^xb_2^x+ib_1^yb_2^y+ic_1c_2+ ib_2^xb_3^x+ib_2^yb_3^y+ic_2c_3\bigg]
}
in the space spanned by $\{\Ket{\vn_1;\vn_2;\vn_3}\}$, we find four states with energy
\eq{
{E_\mathrm{exact}}=-8.4853t.
}

The four-fold degeneracy is lifted when $H_\mathrm{tun}$ is diagonalized in the truncated Hilbert space which includes charge degrees of freedom, $\mathscr{H}_{N_\mathrm{max}}=\{ \Ket{\vn_1;\vn_2;\vn_3}\Ket{\vN}: \max N_i\leq N_\mathrm{max}\}$, and we expect it to reappear as $N_\mathrm{max}$ becomes large.
In Fig.~\ref{fig:CutoffDep}, the resulting energy levels as a function of $N_\mathrm{max}$ are plotted.
As expected, as $N_\mathrm{max}$ is increased, the groundstate energy, $E_0$, and first excited state energy, $E_1$, approach $E_\mathrm{exact}$  (both two-fold degenerate).
There is notably still a gap between $E_\mathrm{exact}$ and both $E_0$ and $E_1$ even for $N_\mathrm{max}=11$. 
However, more importantly, the degeneracy has reasserted itself already by $N_\mathrm{max}=4$.

We will use the (approximate) return of degeneracy to benchmark the size necessary for $N_\mathrm{max}$.
The plots in Fig.~\ref{fig:3IslandVertNumerics} are simulated with $N_\mathrm{max}=7.$
\begin{figure}
\centering
\includegraphics[scale=0.5]{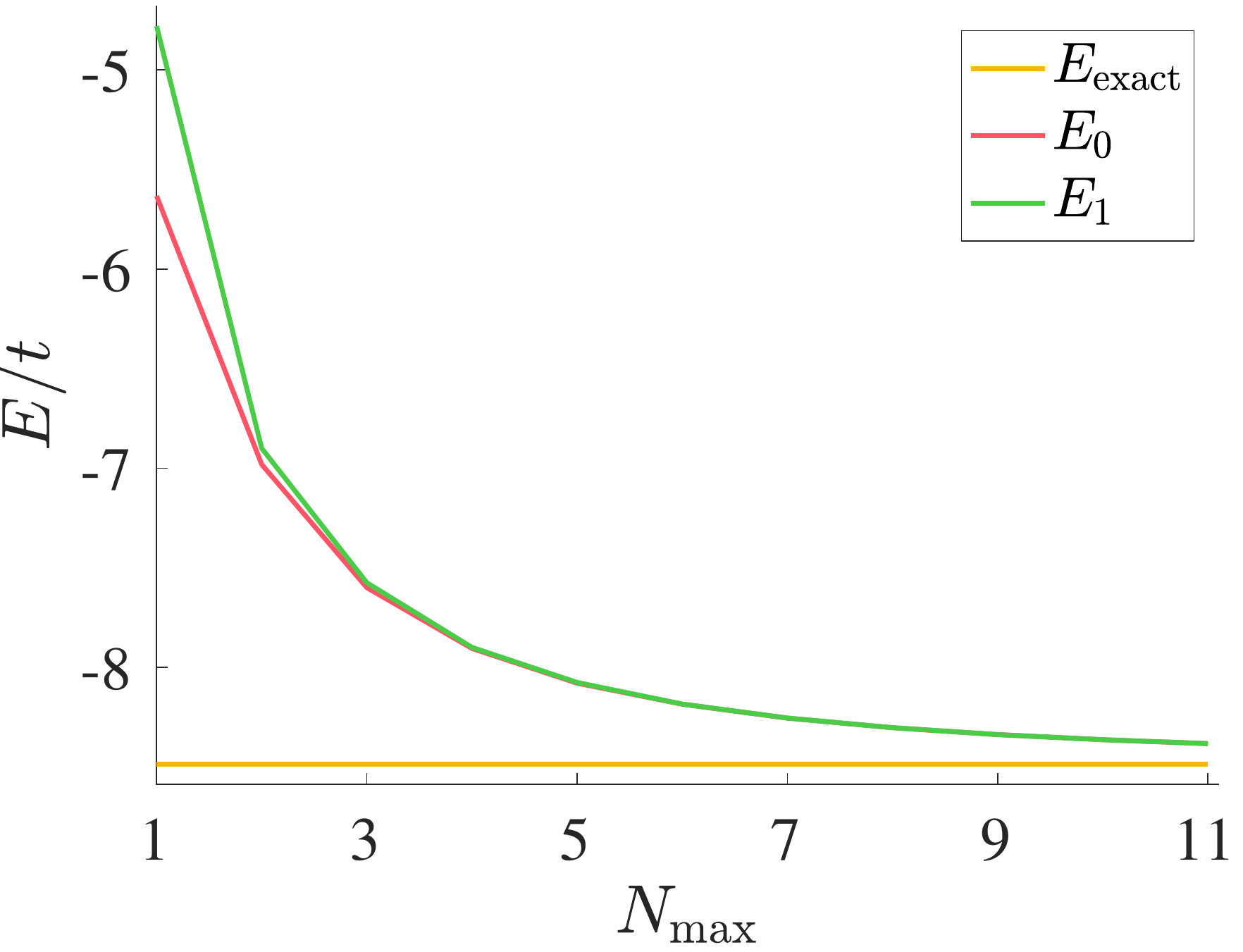}
\caption{The red and green curves plot the lowest and next-lowest groundstates of $H_\mathrm{tun}$ in the Hilbert space $\mathscr{H}_{N_\mathrm{max}}$ for different values of $N_\mathrm{max}$.
The orange curve provides $E_\mathrm{exact}$ as a reference point.
}
\label{fig:CutoffDep}
\end{figure}




}
\end{widetext}



\bibliography{networksRef}
\end{document}